%% file: ee_mod.tex
\documentclass[12pt]{article}
\usepackage{palatino}
\usepackage{axodraw}
\begin{document} 
\vspace*{-1in} 
\renewcommand{\thefootnote}{\fnsymbol{footnote}}
\begin{flushright} 
TIFR/TH/04-12\\
hep-ph/0405292
\end{flushright} 
\vskip 65pt 
\begin{center} 
{\Large \bf \boldmath NLO-QCD corrections to $e^+ e^- \rightarrow$ hadrons in
models of TeV-scale gravity}\\
\vspace{8mm} 
{\bf 
Prakash Mathews${}^{1}$\footnote{pmsp@uohyd.ernet.in}, 
V. Ravindran${}^2$\footnote{ravindra@mri.ernet.in},   
K.~Sridhar${}^3$\footnote{sridhar@theory.tifr.res.in}
}\\ 
\end{center}
\vspace{10pt} 
\begin{flushleft}
{\it 1) School of Physics, University of Hyderabad,
Hyderabad 500 046, India.

2) Harish-Chandra Research Institute, 
 Chhatnag Road, Jhunsi, Allahabad, India.\\

3) Department of Theoretical Physics, 
Tata Institute of Fundamental Research,   
Homi Bhabha Road, Mumbai 400 005, India. } 
\end{flushleft}
 
\vspace{80pt} 
\begin{center}
{\bf ABSTRACT} 
\end{center} 
\vskip12pt 
We present results on NLO-QCD corrections to the process $e^+ e^- \rightarrow$ 
hadrons via photon-, $Z$- and graviton-exchange in the context of TeV-scale 
gravity models. The quantitative impact of these QCD corrections for
searches of extra dimensions at a Linear Collider is briefly discussed.
 
\setcounter{footnote}{0} 
\renewcommand{\thefootnote}{\arabic{footnote}} 
 
\vfill 
\clearpage 
\setcounter{page}{1} 
\pagestyle{plain}
\noindent Models of extra dimensions have dominated the recent 
theoretical literature on physics beyond the standard model.
These models are attractive because one can use
the geometry of the compact extra spatial 
dimensions to stabilise the hierarchy between the electroweak
and the Planck scales.  In the model of Arkani-Hamed, Dimopoulos
and Dvali (the ADD model) \cite{add} a large magnitude of $d$
extra dimensions is responsible for lowering the Planck scale
down to a TeV and the hierarchy problem is thereby avoided. 

In the ADD model, the Standard Model (SM) fields are constrained to a 
3-brane, while gravitons propagate in the $4+d$ dimensions.
Then the size of the extra dimensions is only constrained by the length 
scales to which the gravitational inverse square law has been
experimentally tested, which are currently probing the sub-millimetre range.
For $d$ between 2 and 6, the size of the extra dimensions varies from
a millimetre to a fermi. The relation between the
4-dimensional Planck scale $M_P$ and the scale $M_S$ in $(4+d)$-dimension is 
\begin{equation}
M_P \approx M_S^{(d+2)} R^d ~,
\end{equation}
where $R$ is the compactification radius. In the ADD model, because $R$ is 
large it is possible to lower $M_S$ down to a TeV and avoid the hierarchy
problem. An important consequence of the lowering of the Planck scale
is that quantum gravity effects could be discernible at energies of
${\cal O}$(TeV) and, consequently, a whole new class of studies of
the effects of gravitons at present and future colliders becomes possible.

The graviton propagating in the full $4+d$ dimensions, manifests in 
4-dimensions as a tower of massive 
Kaluza-Klein (KK) modes. These modes interact with the SM particles confined 
to the 3-brane via the energy-momentum tensor.  Each KK mode couples to the SM 
particles with a coupling of the order of $1/M_P$.  Even though 
the coupling of each KK mode to the SM particles are highly 
suppressed.  The effective coupling is obtained after summing over 
all the KK modes and due to the high multiplicity of the KK modes 
the effective interaction has a strength of $1/M_S$ \cite{grw,hlz}.  
This enhanced interaction strength provides viable signatures of
the graviton KK modes at colliders. Both real graviton production
and the effects of virtual gravitons in various processes have
been studied in the literature \cite{rev} and have
yielded bounds on $M_S$ which are in the ball-park of a TeV.

Existing collider studies of the effects of gravitons have
been done at the Born level in QCD. It is important to compute
next-to-leading order QCD (NLO-QCD) to these processes in order to quantify
the size of the QCD correction and to see how robust the leading
order estimate of the cross-section is with respect to the QCD
correction.
Clearly this impacts the experimental studies
of the graviton-mediated processes (and graviton production processes)
crucially. It is with this motivation that the present paper 
presents the results of the computation of NLO-QCD corrections
to $e^+ e^- \rightarrow {\rm hadrons}$ via $\gamma$, Z and graviton 
exchange. These results are used to study the impact of the QCD
correction for this process at the proposed linear $e^+ e^-$ collider
which is planned to operate at centre-of-mass energies between 500
GeV and 1.2 TeV. The results we present here are for the ADD model, but 
since the QCD corrections that we present here are model-independent they
may equally be used for studying the Randall-Sundrum model of warped 
compactification \cite{rs}. 

We work with the following action:
\begin{eqnarray}
{\cal S}={\cal S}_{SM} -{1 \over 2} {\kappa} \int d^4x~
\Theta^{QCD}_{\mu \nu}(x) ~h^{\mu\nu}(x) ~,
\end{eqnarray}
where $S_{SM}$ is the Standard Model action, $h^{\mu\nu}$ is the
graviton field and $\kappa$ is the strength of the interaction.

The energy momentum tensor in QCD is given by
\begin{eqnarray}
\Theta_{\mu \nu}^{QCD}&=&-g_{\mu \nu} {\cal L}_{QCD}
-F_{\mu \rho}^a F_\nu^{a \rho}
-{1 \over \xi}g_{\mu \nu} \partial^\rho
(A_\rho^a \partial^\sigma A_\sigma^a)
\nonumber\\[2ex] &&
+{1 \over \xi}(A_\nu^a \partial_\mu(\partial^\sigma A^a_\sigma)
  +A_\mu^a \partial_\nu(\partial^\sigma A_\sigma^a))
+{i \over 4} \Big[
  \overline \psi \gamma_\mu (\overrightarrow \partial_\nu
-i g T^a A_\nu^a)\psi
\nonumber\\[2ex] &&
 -\overline \psi (\overleftarrow \partial_\nu
+i g T^a A_\nu^a)\gamma_\mu\psi
 +\overline \psi \gamma_\nu (\overrightarrow \partial_\mu
-i g T^a A_\mu^a)\psi
\nonumber\\[2ex] &&
 -\overline \psi (\overleftarrow \partial_\mu
+i g T^a A_\mu^a)\gamma_\nu\psi \Big]
+\partial_\mu \overline \omega^a (\partial_\nu \omega^a
-g f^{abc} A_\nu^c \omega^b)
\nonumber\\[2ex] &&
+\partial_\nu \overline \omega^a (\partial_\mu \omega^a
-g f^{abc} A_\mu^c \omega^b) ~.
\end{eqnarray}
In the above equation, $\xi$ is the gauge fixing parameter.
We work in the Feynman gauge in which the gauge parameter $\xi=1$.
We have displayed explicitly the ghost terms with the ghost fields
$\omega^a(x)$ since they contribute to our one-loop virtual corrections
to the process under study. The presence of the ghost fields introduces
two new vertices viz: \\
1) graviton-ghost-ghost vertex,
\begin{eqnarray}
\Gamma_{\mu \nu}(p_1,p_2)=-i {\kappa \over 2} \delta^{ab}
C_{\mu \nu,\rho \sigma} ~p_1^\rho ~p_2^\sigma  ~,
\end{eqnarray}
2) graviton-ghost-ghost-gluon vertex,
\begin{eqnarray}
\Gamma_{\mu \nu,\rho}(p_1,p_2)&=&-{\kappa \over 2}g f^{abc}
C_{\mu \nu,\rho \sigma} ~p_2^\sigma ~,
\end{eqnarray}
where
\begin{eqnarray}
C_{\mu \nu,\rho \sigma} &=& g_{\mu\rho} g_{\nu\sigma} 
                          + g_{\mu\sigma} g_{\nu\rho}
                          - g_{\mu\nu}  g_{\rho\sigma} ~,
\nonumber
\end{eqnarray}
the $\mu,\nu$ index refers to the graviton field and $p_1, ~b$ and $p_2, 
~a$ are the momenta and colour index of the incoming and out going 
ghost field respectively.  Other than the rules for these vertices, our 
Feynman rules are the same as that given in Ref.~\cite{hlz} \footnote
{The only exceptions are the Feynman rules for the fermion-antifermion-gauge 
boson-graviton vertex and the three gauge boson-graviton vertex which differ
from those of Ref.~\cite{hlz} by an overall sign.}

We start by considering $e^+~e^-$ scattering to hadronic final states,  
\begin{eqnarray}
l^-(k_1)+l^+(k_2) \rightarrow X(P_X)  ~,
\end{eqnarray}
where $l^-,l^+$ are the incoming leptons and $X$ are the final-state hadrons 
and $k_1,k_2$ and  $P_X$ are their respective momenta.
The cross-section can be factorised into a leptonic 
part ${\cal L}_i(k_1,k_2)$ 
and a hadronic part ${\cal W}_i(q)$, as follows: 
\begin{eqnarray}
\sigma^{e^+~e^-} (k_1,k_2)&=&
{1 \over 2 s} 
\sum_{i=\gamma, Z, G} \int {d^n q \over (2 \pi)^n} (2 \pi)^n 
\delta^{(n)}(q-k_1-k_2) 
\nonumber\\[2ex]
&&{\cal L}_i(k_1,k_2)\cdot
{\cal P}_i(q)\cdot{\cal P}_i(q)\cdot~{\cal W}_i(q) ~,
\end{eqnarray}
where $n$ is the space-time dimension, $q$ is the momentum of the intermediate 
photon, $Z$ or graviton and $\sqrt{s}$ is the center of mass energy, 
\begin{eqnarray}
s&=&(k_1+k_2)^2 ~,
\nonumber\\[2ex]
&=&q^2=Q^2 ~.
\end{eqnarray}
The propagator ${\cal P}_G(q)$ is given by
\begin{eqnarray}
{\cal P}_G(q)=i ~B_{\mu \nu \rho \sigma}(q) {\cal D}(q^2) ~,
\end{eqnarray}
where,
\begin{eqnarray}
B_{\mu \nu \rho \sigma}(q)=
\eta_{\mu \rho} \eta_{\nu \sigma}
+\eta_{\mu \sigma} \eta_{\nu \rho}
-{2 \over 3} \eta_{\mu \nu} \eta_{\rho \sigma} ~,
\nonumber 
\end{eqnarray}
with
\begin{eqnarray}
\eta_{\mu \nu}&=&-g_{\mu \nu} + {q_\mu q_\nu \over Q^2} ~,
\nonumber
\end{eqnarray}
The function ${\cal D}(q^2)$ results from the sum over the 
time-like propagators of the KK modes:
\begin{eqnarray}
{\cal D}(q^2)=16 \pi~ \left({Q^{d-2} \over \kappa^2 M_S^{d+2}}\right)~
I\left({M_S \over Q}\right) ~,
\end{eqnarray}
the summation over the non-resonant KK modes yields
\begin{eqnarray}
I(\omega)&=&- \sum_{k=1}^{d/2-1} {1 \over 2 k} \omega^{2 k}
 -{1 \over 2} \log(\omega^2-1) \qquad \qquad \qquad  d={\rm even},
\nonumber\\[2ex]
&=&
-\sum_{k=1}^{(d-1)/2} {1\over 2 k-1} \omega^{2 k-1}
 +{1 \over 2} \log \left({\omega+1}\over{\omega-1}\right) 
\quad \quad  d={\rm odd},
\end{eqnarray}
where $\omega=M_S/Q$.

The leptonic tensor involves just the computation of square of
the matrix element for the process $e^+ + e^- \rightarrow \gamma^*/Z^*(q)$ 
and $e^+ + e^- \rightarrow G^*(q)$ and is given as
\begin{eqnarray}
{\cal L}_i(k_1,k_2)&=&
{1 \over 4} \sum_{spin} |M^{e^+~e^- \rightarrow i}|^2
 \quad \quad i=\gamma, Z, G.
\end{eqnarray}

The hadronic part ${\cal W}_i(q)$ is computed using 
\begin{eqnarray}
{\cal W}_i(q)&=&\int \prod_{j=1}^m \left({d^n p_j \over (2 \pi)^n}
(2 \pi) \delta^+(p_j^2)\right)
\nonumber \\
&& \times (2 \pi)^n \delta^n\Big(q+\sum_{j=1}^m p_j\Big) 
|M^{i \rightarrow \sum_{j=1}^m p_j}|^2  \quad \quad i=\gamma, Z, G.
\end{eqnarray}
\begin{figure}[hbt]
\SetScale{0.8}
\noindent
\begin{picture}(150,120)(-20,-10)
\DashArrowLine(50,50)(100,50){4} \Text(60,55)[]{$G^*$}
\Gluon(100,50)(150,100){3}{4}
\ArrowLine(150,100)(150,0)
\Gluon(150,0)(100,50){3}{4}
\ArrowLine(200,100)(150,100) \Text(170,80)[]{$\bar q$}
\ArrowLine(150,0)(200,0)\Text(170,0)[]{$q$}
\end{picture}
\hspace*{1em}
\begin{picture}(150,120)(-20,-10)
\DashArrowLine(50,50)(100,50){4} \Text(60,55)[]{$G^*$}
\ArrowLine(150,75)(100,50)
\ArrowLine(200,100)(150,75) \Text(170,80)[]{$\bar q$}
\Gluon(150,75)(150,25){3}{4}
\ArrowLine(150,25)(200,0) \Text(170,0)[]{$q$}
\ArrowLine(100,50)(150,25)
\end{picture} \\{\sl }

\caption{\sl Graviton-quark-antiquark vertex at one loop.}
\label{virtq}
\end{figure}
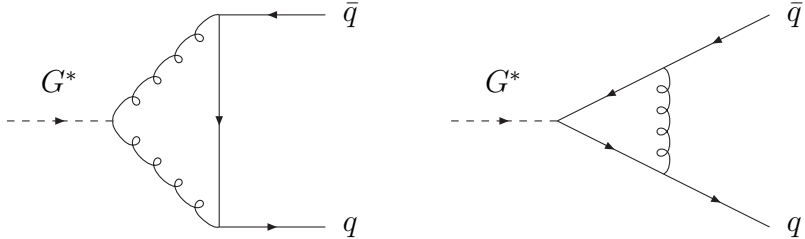

In the following, we compute leading (LO) and next-to-leading order(NLO)
contributions to the hadronic part within perturbative QCD
in powers of the strong-coupling constant $\alpha_s$. 
Since we have gravity-mediated process, in addition to the usual SM
process, we have two classes of diagrams.
The first class contains only photon or $Z$ in the intermediate state
and the second class contains only gravitons. The interference between
SM and gravity channels is identically zero in the fully inclusive 
reaction.  
Notice that the interference term will always be proportional to
the product of a third-rank leptonic tensor and a third-rank hadronic
tensor contracted via the propagator tensors of either $\gamma$ or $Z$
with that of the graviton.  For the 
photon-graviton case, the inclusive hadronic tensor can be written 
only in terms of $g_{\mu \nu}$ and $q_\rho$ (the Levi-Cevita tensor 
will not appear in this).  We know that there is no third rank 
tensor, say $S^{\mu \nu \rho}$ which can be constructed out of 
$g_{\mu \nu}$ and $q_\rho$ ($q.q\not=0$) that satisfies $q_\mu S^{\mu 
\nu \rho}=0$ (as it should be for the theory where we have gravity 
coupled to a conserved energy momentum tensor).  In the case of 
Z-graviton interference, the hadronic tensor can be proportional to a 
tensor, say, $S^{\mu \nu \rho} =\epsilon^{\mu \nu \rho \sigma} 
q_\sigma$ which satisfies $q_\mu S^{\mu \nu \rho}=0$, but this when 
contracted with the Graviton propagator vanishes identically.  Hence, 
there is no graviton-SM interference contribution to inclusive
s-channel cross section (one can also argue out based on
angular momentum conservation in field theories).
\begin{figure}[hbt]
\SetScale{0.8}
\noindent
\begin{picture}(150,120)(-20,-10)
\DashArrowLine(28,50)(78,50){4}\Text(40,55)[]{$G^*$}
\GlueArc(100,50)(20,0,180){3}{3}
\GlueArc(100,50)(20,180,360){3}{3}
\Gluon(120,50)(180,90){5}{4} \Text(155,75)[]{$g$}
\Gluon(120,50)(180,10){-5}{4}\Text(155,10)[]{$g$}
\end{picture}
\hspace*{1em}
\begin{picture}(150,120)(-20,-10)
\DashArrowLine(50,50)(100,50){4}\Text(60,55)[]{$G^*$}
\Gluon(100,50)(150,100){3}{5}
\Gluon(150,100)(150,0){3}{5}
\Gluon(150,0)(100,50){3}{5}
\Gluon(150,100)(200,100){5}{4}\Text(170,80)[]{$g$}
\Gluon(150,0)(200,0){-5}{4}\Text(170,0)[]{$g$}
\end{picture} \\{\sl }
\vspace*{.8em}
\begin{picture}(150,120)(-20,-10)
\DashArrowLine(50,50)(100,50){4}\Text(60,55)[]{$G^*$}
\ArrowLine(100,50)(150,100)
\ArrowLine(150,100)(150,0)
\ArrowLine(150,0)(100,50)
\Gluon(150,100)(200,100){5}{4}\Text(170,80)[]{$g$}
\Gluon(150,0)(200,0){-5}{4}\Text(170,0)[]{$g$}
\end{picture}
\hspace*{.5em}
\begin{picture}(150,120)(-20,-10)
\DashArrowLine(50,50)(100,50){4}\Text(60,55)[]{$G^*$}
\ArrowLine(150,100)(100,50)
\ArrowLine(150,0)(150,100)
\ArrowLine(100,50)(150,0)
\Gluon(150,100)(200,100){5}{4}\Text(170,80)[]{$g$}
\Gluon(150,0)(200,0){-5}{4}\Text(170,0)[]{$g$}
\end{picture} \\ {\centerline{\sl + ghost loops}}
\caption{\sl Graviton-gluon-gluon vertex at one loop.}
\label{virtg}
\end{figure}
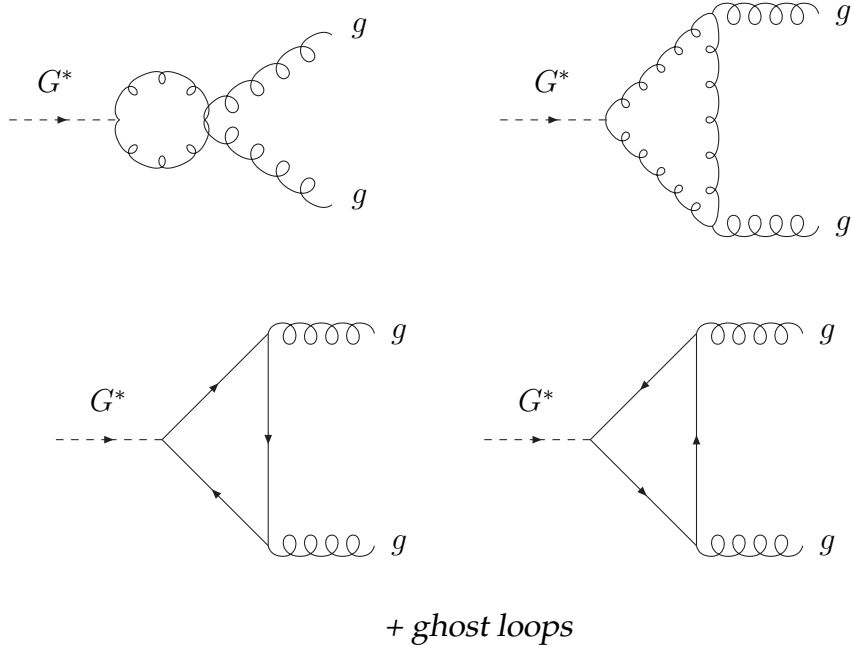

The leading-order contribution to the hadronic
part comes from
\begin{eqnarray}
\gamma^*/Z^* &\rightarrow& q~+~\bar q ~,
\end{eqnarray}
and at next to leading order level we have
\begin{eqnarray}
\gamma^*/Z^* &\rightarrow& q~+~\bar q + {\rm one~loop},
\nonumber\\[2ex]
\gamma^*/Z^* &\rightarrow& q~+~\bar q + g ~.
\end{eqnarray}
\begin{figure}[hbt]
\SetScale{0.8}
\noindent
\begin{picture}(150,120)(-20,-10)
\DashArrowLine(50,50)(100,50){4}\Text(60,55)[]{$G^*$}
\ArrowLine(150,75)(100,50) \Text(170,75)[]{$\bar q$}
\ArrowLine(200,100)(150,75)
\Gluon(150,75)(200,50){5}{4} \Text(170,45)[]{$g$}
\ArrowLine(100,50)(200,0)\Text(170,0)[]{$ q$}
\end{picture}
\hspace*{2em}
\begin{picture}(150,120)(-20,-10)
\DashArrowLine(50,50)(100,50){4}\Text(60,55)[]{$G^*$}
\ArrowLine(200,100)(100,50)\Text(170,75)[]{$\bar q$}
\Gluon(150,25)(200,50){5}{4}\Text(170,45)[]{$g$}
\ArrowLine(150,25)(200,0)\Text(170,0)[]{$ q$}
\ArrowLine(100,50)(150,25)
\end{picture}
\end{figure}
\begin{figure}[hbt]
\SetScale{0.8}
\noindent
\begin{picture}(150,120)(-20,-10)
\DashArrowLine(50,50)(100,50){4}\Text(60,55)[]{$G^*$}
\ArrowLine(200,100)(100,50)\Text(170,75)[]{$\bar q$}
\Gluon(100,50)(200,50){4}{8}\Text(170,45)[]{$g$}
\ArrowLine(100,50)(200,0)\Text(170,0)[]{$ q$}
\end{picture}
\hspace*{2em}
\begin{picture}(150,120)(-20,-10)
\DashArrowLine(50,50)(100,50){4}\Text(60,55)[]{$G^*$}
\Gluon(100,50)(150,75){4}{4}
\ArrowLine(200,100)(150,75)\Text(170,75)[]{$\bar q$}
\ArrowLine(150,75)(200,50)\Text(170,45)[]{$q$}
\Gluon(200,0)(100,50){4}{8}\Text(170,0)[]{$ g$}
\end{picture}
\caption{$G^* \rightarrow q +\bar q + g$.}
\label{realq}
\end{figure}
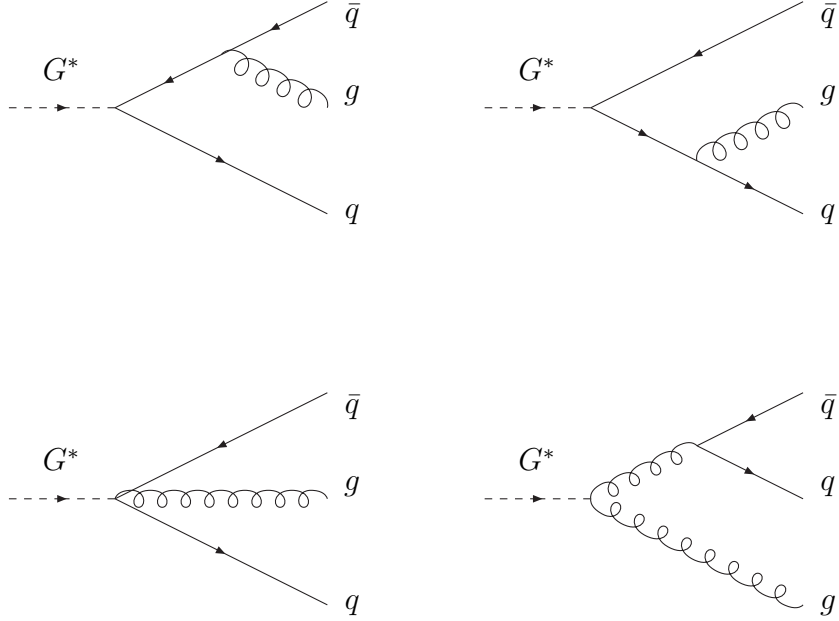

When the graviton is exchanged, we have two tree-level contributions
\begin{eqnarray}
G^* &\rightarrow& q~+~\bar q ~,
\nonumber\\[2ex]
G^* &\rightarrow& g~+~ g ~.
\end{eqnarray}
At NLO, the following processes, as shown in Fig.~(1 -- 4), contribute 
\begin{eqnarray}
G^* &\rightarrow& q~+~\bar q +{\rm one~loop},
\nonumber\\[2ex]
G^* &\rightarrow& q~+~\bar q +g ~,
\nonumber\\[2ex]
G^* &\rightarrow& g~+~g +{\rm one~loop},
\nonumber\\[2ex]
G^* &\rightarrow& g~+~g + g ~.
\end{eqnarray}
\begin{figure}
\SetScale{0.8}
\noindent
\begin{picture}(150,120)(-20,-10)
\DashArrowLine(50,50)(100,50){4}\Text(60,55)[]{$G^*$}
\Gluon(100,50)(200,100){4}{8} \Text(180,80)[]{$g(p_1)$}
\Gluon(200,0)(100,50){4}{8} \Text(180,0)[]{$g(p_2)$}
\Gluon(100,50)(200,50){4}{8}\Text(180,45)[]{$g(p_3)$}
\end{picture}
\hspace*{2em}
\begin{picture}(150,120)(-20,-10)
\DashArrowLine(50,50)(100,50){4}\Text(60,55)[]{$G^*$}
\Gluon(100,50)(150,75){4}{4} \Text(180,80)[]{$g(p_1)$}
\Gluon(150,75)(200,100){4}{4}
\Gluon(200,50)(150,75){4}{4}\Text(180,0)[]{$g(p_2)$}
\Gluon(200,0)(100,50){4}{8}\Text(180,40)[]{$g(p_3)$}
\label{realg}
\end{picture}
\\\centerline{\sl + permutations}
\caption{$G^* \rightarrow g+g+g$.}
\end{figure}
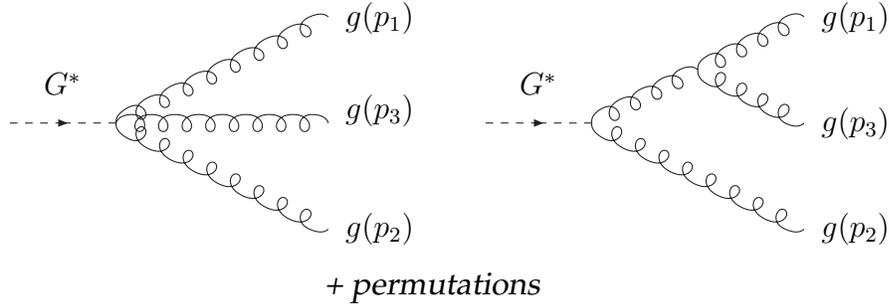
The next-to-leading order corrections involve the computation of
one-loop virtual gluon corrections and real gluon bremsstrahlung contributions
to leading-order processes.  Since we are dealing with the energy-momentum
tensor coupled to gravity expressed in terms of renormalised fields and
massless quarks,  there is no overall ultraviolet renormalisation required.
In other words, the operator renormalisation constant for the
energy-momentum operator is identical to unity to all orders in perturbation
theory.  But we encounter infrared divergences (soft and collinear) 
in our computation.  We have used dimensional regularisation to
regulate these divergences.  To do this, we have defined $n=4+\varepsilon$
where $n$ is space-time dimension.  With this, all these divergences
appear as $1/\varepsilon^\alpha$ where $\alpha=1,2$.
Since the gravitons couple directly to quarks and gluons,
there are many one-loop diagrams which appear to this order. But most
of them do not contribute when quarks are taken to be massless.
The contributing diagrams are given in the Figs.~(1,2).
Since throughout our calculation we have used the Feynman gauge,
the gravitons also couple to ghost fields which can appear in loops.  
The real gluon-emission diagrams are given in the 
Figs.~(3,4).  In the Fig.~(4), three identical gluons in the final state     
are permuted with the appropriate statistical factor.
The soft divergences coming from
virtual gluons and bremsstrahlung contributions cancel exactly as expected.
Similarly, the remaining collinear divergences disappear in the fully inclusive
reaction.

The hadronic part can be expanded in powers of 
\begin{eqnarray}
a_s={\alpha_s \over 4 \pi} ~,
\end{eqnarray}
as follows:
\begin{eqnarray}
{\cal W}_i={\cal W}_i^{(0)}+ a_s~ {\cal W}_i^{(1)}  
\quad \quad i=\gamma, Z, G.
\end{eqnarray}
After adding all the diagrams and folding in the appropriate colour factors 
we arrive at a simple-looking result for ${\cal W}_i^{(0)}$ and 
${\cal W}_i^{(1)}$ for the non-SM part

\begin{eqnarray}
   {\cal W}^{(0)}_{G,\mu \nu \rho \sigma} &=&
        {Q^4  \kappa^2\over \pi}~\Bigg[ (N^2-1) \left({1 \over 320}\right)
                                   +N~n_f \left({1 \over 640}\right)\Bigg]
        B_{\mu\nu\rho\sigma}(q) ~,
\nonumber\\[2ex]
   {\cal W}^{(1)}_{G,\mu \nu \rho \sigma} &=&
        a_s~ {Q^4  \kappa^2\over \pi}
        ~\Bigg[ (N^2-1) C_A\left(-{1 \over 144}\right)
                +N n_f C_F \left({7 \over 1152}\right)\Bigg]
        B_{\mu\nu\rho\sigma}(q) ~.
\end{eqnarray}
The colour factors appearing
in the above equations are
\begin{eqnarray}
C_F={N^2 -1 \over 2 N} ~, \quad \quad C_A=N ~,
\end{eqnarray}
for $SU(N)$ gauge theory and $n_f$ is the number of flavours.
The total cross section is found to be
\begin{eqnarray}
       \sigma^{SM}&=& \sum_{q} ~{4 \pi ~\alpha^2 \over 3 ~s} N~
\left\{ Q_q^2- Q_q ~{2 s\over c_w^2 s_w^2} ~g^V_e ~g^V_q {(s-M_Z^2) 
\over (s-M_Z^2)^2+M_Z^2 \Gamma_Z^2} \right.
\nonumber \\[2ex]
&&
\left.
+{s^2 \over c_w^4 s_w^4} ({g^V_e}^2+{g^A_e}^2)  ({g^V_q}^2+{g^A_q}^2) 
{1 \over {(s-M_Z^2)^2 + M_Z^2 \Gamma_Z^2}} \right\}
             \Bigg(1+ a_s~(3 C_F)\Bigg) ~,
\nonumber \\[2ex]
    \sigma^{G}&=& {\kappa^4 \over 2560 \pi s}Q^8 
    \left[{\cal D}(q^2)\right]^2
        \Bigg\lbrace (N^2-1) \left(1 + a_s~\left(-{20 \over 9}\right)C_A\right)
\nonumber\\[2ex]
&&   + {N n_f \over 2} \left(1 + a_s~{35 \over 9}C_F\right)\Bigg\rbrace~,
\end{eqnarray}
where
\begin{eqnarray}
c_w &=& \sin \theta_W ~, \qquad \qquad s_w=\sin \theta_W ~,
\nonumber \\[2ex]
g_a^V &=& {1 \over 2 } T^3_a - s_w^2 Q_a ~, 
\qquad
g_a^A=-{1 \over 2 } T^3_a ~,
\nonumber 
\end{eqnarray}
and $Q_a$ is the electric charge of quarks and leptons.
\begin{figure}[ht]
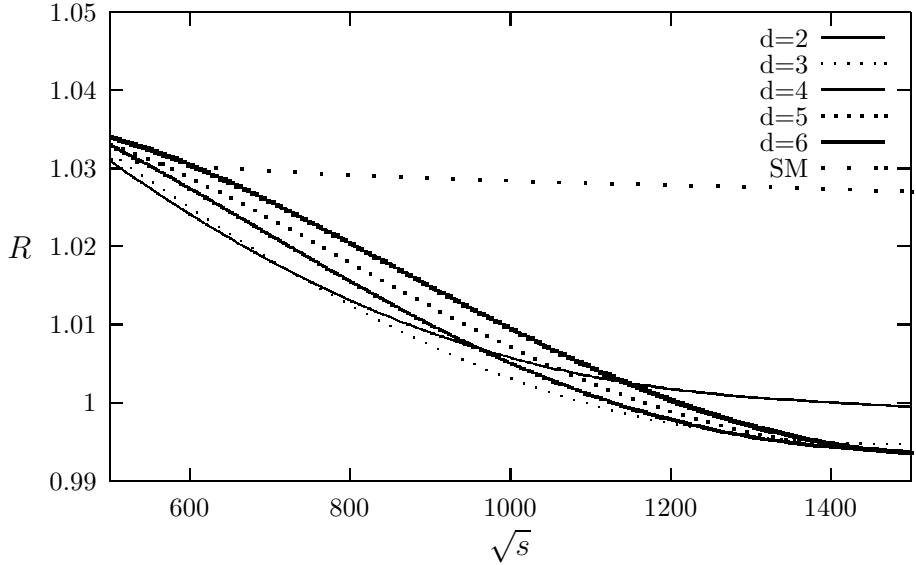

\begin{center}
\input fig1.tex
\end{center}
\vspace*{-20pt}
\caption{\footnotesize\it The ratio $R$ of the LO+NLO cross-section
to the LO cross-section as a function of the $e^+ e^-$ centre-of-mass
energy, for number of extra dimensions, $d$, between 2 and 6. The
SM reference curve is also shown.} 
\end{figure}
We now proceed to estimate the magnitude of the QCD corrections presented 
above for the process $e^+ e^- \rightarrow {\rm hadrons}$ at the Linear
Collider. In Fig.~(5), we have plotted the ratio, $R$ as a function of
the linear collider centre-of-mass energy, where 
\begin{equation}
R={\sigma_{\rm LO+NLO} \over \sigma_{\rm LO}} ~.
\end{equation}
The figure shows $R$ as a function of $\sqrt{s}$ for different values
of the number of extra dimensions, $d$. Also shown in the figure is the
ratio $R$ for the SM case. For our estimates, we have used $\alpha_s$
evaluated at the scale $s$ and for the running of $\alpha_s$ we have
used $n_f=5$ and $\Lambda=0.215$~GeV. While the corrections are typically 
of the order of a couple of percent, they could still be significant
because of the large statistics for this
process at the Linear Collider (even if one assumes
a conservative $50 - 100$ fb${}^{-1}$ luminosity). Because of the
the gluon-graviton couplings, 
the corrections coming from the graviton-exchange diagrams give a
negative contribution to the NLO corrections, which tends to drive
$R$ to values below 1 at large centre-of-mass energies. 

In conclusion, we have presented the first computation of NLO-QCD corrections
to graviton-mediated processes in the context of models of large extra 
dimensions. The process that we have considered is $e^+ e^- \rightarrow
{\rm hadrons}$ via $\gamma,\ Z,\ {\rm and\ graviton}$ exchange. We have
discussed the impact of the QCD corrections for the study of the ADD model
using this process at a Linear Collider. 

\vspace{20pt}
\noindent
{\it Acknowledgments:} The work of Prakash Mathews and K. Sridhar is
part of a project (IFCPAR Project No. 2904-2) supported by the 
Indo-French Centre for the Promotion of Advanced Research, New Delhi, India.
They would also like to thank the CERN Theory Division for hospitality
when this work was being completed.
 

\end{document}

%% file: fig1.tex
\setlength{\unitlength}{0.240900pt}
\ifx\plotpoint\undefined\newsavebox{\plotpoint}\fi
\sbox{\plotpoint}{\rule[-0.200pt]{0.400pt}{0.400pt}}%
\begin{picture}(1500,900)(0,0)
\font\gnuplot=cmr10 at 10pt
\gnuplot
\sbox{\plotpoint}{\rule[-0.200pt]{0.400pt}{0.400pt}}%
\put(181.0,123.0){\rule[-0.200pt]{4.818pt}{0.400pt}}
\put(161,123){\makebox(0,0)[r]{0.99}}
\put(1419.0,123.0){\rule[-0.200pt]{4.818pt}{0.400pt}}
\put(181.0,246.0){\rule[-0.200pt]{4.818pt}{0.400pt}}
\put(161,246){\makebox(0,0)[r]{1}}
\put(1419.0,246.0){\rule[-0.200pt]{4.818pt}{0.400pt}}
\put(181.0,369.0){\rule[-0.200pt]{4.818pt}{0.400pt}}
\put(161,369){\makebox(0,0)[r]{1.01}}
\put(1419.0,369.0){\rule[-0.200pt]{4.818pt}{0.400pt}}
\put(181.0,492.0){\rule[-0.200pt]{4.818pt}{0.400pt}}
\put(161,492){\makebox(0,0)[r]{1.02}}
\put(1419.0,492.0){\rule[-0.200pt]{4.818pt}{0.400pt}}
\put(181.0,614.0){\rule[-0.200pt]{4.818pt}{0.400pt}}
\put(161,614){\makebox(0,0)[r]{1.03}}
\put(1419.0,614.0){\rule[-0.200pt]{4.818pt}{0.400pt}}
\put(181.0,737.0){\rule[-0.200pt]{4.818pt}{0.400pt}}
\put(161,737){\makebox(0,0)[r]{1.04}}
\put(1419.0,737.0){\rule[-0.200pt]{4.818pt}{0.400pt}}
\put(181.0,860.0){\rule[-0.200pt]{4.818pt}{0.400pt}}
\put(161,860){\makebox(0,0)[r]{1.05}}
\put(1419.0,860.0){\rule[-0.200pt]{4.818pt}{0.400pt}}
\put(307.0,123.0){\rule[-0.200pt]{0.400pt}{4.818pt}}
\put(307,82){\makebox(0,0){600}}
\put(307.0,840.0){\rule[-0.200pt]{0.400pt}{4.818pt}}
\put(558.0,123.0){\rule[-0.200pt]{0.400pt}{4.818pt}}
\put(558,82){\makebox(0,0){800}}
\put(558.0,840.0){\rule[-0.200pt]{0.400pt}{4.818pt}}
\put(810.0,123.0){\rule[-0.200pt]{0.400pt}{4.818pt}}
\put(810,82){\makebox(0,0){1000}}
\put(810.0,840.0){\rule[-0.200pt]{0.400pt}{4.818pt}}
\put(1062.0,123.0){\rule[-0.200pt]{0.400pt}{4.818pt}}
\put(1062,82){\makebox(0,0){1200}}
\put(1062.0,840.0){\rule[-0.200pt]{0.400pt}{4.818pt}}
\put(1313.0,123.0){\rule[-0.200pt]{0.400pt}{4.818pt}}
\put(1313,82){\makebox(0,0){1400}}
\put(1313.0,840.0){\rule[-0.200pt]{0.400pt}{4.818pt}}
\put(181.0,123.0){\rule[-0.200pt]{303.052pt}{0.400pt}}
\put(1439.0,123.0){\rule[-0.200pt]{0.400pt}{177.543pt}}
\put(181.0,860.0){\rule[-0.200pt]{303.052pt}{0.400pt}}
\put(40,491){\makebox(0,0){$R$}}
\put(810,21){\makebox(0,0){$\sqrt{s}$}}
\put(181.0,123.0){\rule[-0.200pt]{0.400pt}{177.543pt}}
\put(1279,820){\makebox(0,0)[r]{d=2}}
\put(1299.0,820.0){\rule[-0.200pt]{24.090pt}{0.400pt}}
\put(181,627){\usebox{\plotpoint}}
\multiput(181.00,625.93)(0.728,-0.489){15}{\rule{0.678pt}{0.118pt}}
\multiput(181.00,626.17)(11.593,-9.000){2}{\rule{0.339pt}{0.400pt}}
\multiput(194.00,616.93)(0.669,-0.489){15}{\rule{0.633pt}{0.118pt}}
\multiput(194.00,617.17)(10.685,-9.000){2}{\rule{0.317pt}{0.400pt}}
\multiput(206.00,607.93)(0.728,-0.489){15}{\rule{0.678pt}{0.118pt}}
\multiput(206.00,608.17)(11.593,-9.000){2}{\rule{0.339pt}{0.400pt}}
\multiput(219.00,598.93)(0.824,-0.488){13}{\rule{0.750pt}{0.117pt}}
\multiput(219.00,599.17)(11.443,-8.000){2}{\rule{0.375pt}{0.400pt}}
\multiput(232.00,590.93)(0.728,-0.489){15}{\rule{0.678pt}{0.118pt}}
\multiput(232.00,591.17)(11.593,-9.000){2}{\rule{0.339pt}{0.400pt}}
\multiput(245.00,581.93)(0.758,-0.488){13}{\rule{0.700pt}{0.117pt}}
\multiput(245.00,582.17)(10.547,-8.000){2}{\rule{0.350pt}{0.400pt}}
\multiput(257.00,573.93)(0.728,-0.489){15}{\rule{0.678pt}{0.118pt}}
\multiput(257.00,574.17)(11.593,-9.000){2}{\rule{0.339pt}{0.400pt}}
\multiput(270.00,564.93)(0.824,-0.488){13}{\rule{0.750pt}{0.117pt}}
\multiput(270.00,565.17)(11.443,-8.000){2}{\rule{0.375pt}{0.400pt}}
\multiput(283.00,556.93)(0.758,-0.488){13}{\rule{0.700pt}{0.117pt}}
\multiput(283.00,557.17)(10.547,-8.000){2}{\rule{0.350pt}{0.400pt}}
\multiput(295.00,548.93)(0.824,-0.488){13}{\rule{0.750pt}{0.117pt}}
\multiput(295.00,549.17)(11.443,-8.000){2}{\rule{0.375pt}{0.400pt}}
\multiput(308.00,540.93)(0.824,-0.488){13}{\rule{0.750pt}{0.117pt}}
\multiput(308.00,541.17)(11.443,-8.000){2}{\rule{0.375pt}{0.400pt}}
\multiput(321.00,532.93)(0.758,-0.488){13}{\rule{0.700pt}{0.117pt}}
\multiput(321.00,533.17)(10.547,-8.000){2}{\rule{0.350pt}{0.400pt}}
\multiput(333.00,524.93)(0.950,-0.485){11}{\rule{0.843pt}{0.117pt}}
\multiput(333.00,525.17)(11.251,-7.000){2}{\rule{0.421pt}{0.400pt}}
\multiput(346.00,517.93)(0.824,-0.488){13}{\rule{0.750pt}{0.117pt}}
\multiput(346.00,518.17)(11.443,-8.000){2}{\rule{0.375pt}{0.400pt}}
\multiput(359.00,509.93)(0.950,-0.485){11}{\rule{0.843pt}{0.117pt}}
\multiput(359.00,510.17)(11.251,-7.000){2}{\rule{0.421pt}{0.400pt}}
\multiput(372.00,502.93)(0.758,-0.488){13}{\rule{0.700pt}{0.117pt}}
\multiput(372.00,503.17)(10.547,-8.000){2}{\rule{0.350pt}{0.400pt}}
\multiput(384.00,494.93)(0.950,-0.485){11}{\rule{0.843pt}{0.117pt}}
\multiput(384.00,495.17)(11.251,-7.000){2}{\rule{0.421pt}{0.400pt}}
\multiput(397.00,487.93)(0.950,-0.485){11}{\rule{0.843pt}{0.117pt}}
\multiput(397.00,488.17)(11.251,-7.000){2}{\rule{0.421pt}{0.400pt}}
\multiput(410.00,480.93)(0.874,-0.485){11}{\rule{0.786pt}{0.117pt}}
\multiput(410.00,481.17)(10.369,-7.000){2}{\rule{0.393pt}{0.400pt}}
\multiput(422.00,473.93)(0.950,-0.485){11}{\rule{0.843pt}{0.117pt}}
\multiput(422.00,474.17)(11.251,-7.000){2}{\rule{0.421pt}{0.400pt}}
\multiput(435.00,466.93)(0.950,-0.485){11}{\rule{0.843pt}{0.117pt}}
\multiput(435.00,467.17)(11.251,-7.000){2}{\rule{0.421pt}{0.400pt}}
\multiput(448.00,459.93)(1.123,-0.482){9}{\rule{0.967pt}{0.116pt}}
\multiput(448.00,460.17)(10.994,-6.000){2}{\rule{0.483pt}{0.400pt}}
\multiput(461.00,453.93)(0.874,-0.485){11}{\rule{0.786pt}{0.117pt}}
\multiput(461.00,454.17)(10.369,-7.000){2}{\rule{0.393pt}{0.400pt}}
\multiput(473.00,446.93)(1.123,-0.482){9}{\rule{0.967pt}{0.116pt}}
\multiput(473.00,447.17)(10.994,-6.000){2}{\rule{0.483pt}{0.400pt}}
\multiput(486.00,440.93)(0.950,-0.485){11}{\rule{0.843pt}{0.117pt}}
\multiput(486.00,441.17)(11.251,-7.000){2}{\rule{0.421pt}{0.400pt}}
\multiput(499.00,433.93)(1.033,-0.482){9}{\rule{0.900pt}{0.116pt}}
\multiput(499.00,434.17)(10.132,-6.000){2}{\rule{0.450pt}{0.400pt}}
\multiput(511.00,427.93)(1.123,-0.482){9}{\rule{0.967pt}{0.116pt}}
\multiput(511.00,428.17)(10.994,-6.000){2}{\rule{0.483pt}{0.400pt}}
\multiput(524.00,421.93)(1.123,-0.482){9}{\rule{0.967pt}{0.116pt}}
\multiput(524.00,422.17)(10.994,-6.000){2}{\rule{0.483pt}{0.400pt}}
\multiput(537.00,415.93)(1.123,-0.482){9}{\rule{0.967pt}{0.116pt}}
\multiput(537.00,416.17)(10.994,-6.000){2}{\rule{0.483pt}{0.400pt}}
\multiput(550.00,409.93)(1.033,-0.482){9}{\rule{0.900pt}{0.116pt}}
\multiput(550.00,410.17)(10.132,-6.000){2}{\rule{0.450pt}{0.400pt}}
\multiput(562.00,403.93)(1.378,-0.477){7}{\rule{1.140pt}{0.115pt}}
\multiput(562.00,404.17)(10.634,-5.000){2}{\rule{0.570pt}{0.400pt}}
\multiput(575.00,398.93)(1.123,-0.482){9}{\rule{0.967pt}{0.116pt}}
\multiput(575.00,399.17)(10.994,-6.000){2}{\rule{0.483pt}{0.400pt}}
\multiput(588.00,392.93)(1.267,-0.477){7}{\rule{1.060pt}{0.115pt}}
\multiput(588.00,393.17)(9.800,-5.000){2}{\rule{0.530pt}{0.400pt}}
\multiput(600.00,387.93)(1.378,-0.477){7}{\rule{1.140pt}{0.115pt}}
\multiput(600.00,388.17)(10.634,-5.000){2}{\rule{0.570pt}{0.400pt}}
\multiput(613.00,382.93)(1.378,-0.477){7}{\rule{1.140pt}{0.115pt}}
\multiput(613.00,383.17)(10.634,-5.000){2}{\rule{0.570pt}{0.400pt}}
\multiput(626.00,377.93)(1.033,-0.482){9}{\rule{0.900pt}{0.116pt}}
\multiput(626.00,378.17)(10.132,-6.000){2}{\rule{0.450pt}{0.400pt}}
\multiput(638.00,371.94)(1.797,-0.468){5}{\rule{1.400pt}{0.113pt}}
\multiput(638.00,372.17)(10.094,-4.000){2}{\rule{0.700pt}{0.400pt}}
\multiput(651.00,367.93)(1.378,-0.477){7}{\rule{1.140pt}{0.115pt}}
\multiput(651.00,368.17)(10.634,-5.000){2}{\rule{0.570pt}{0.400pt}}
\multiput(664.00,362.93)(1.378,-0.477){7}{\rule{1.140pt}{0.115pt}}
\multiput(664.00,363.17)(10.634,-5.000){2}{\rule{0.570pt}{0.400pt}}
\multiput(677.00,357.94)(1.651,-0.468){5}{\rule{1.300pt}{0.113pt}}
\multiput(677.00,358.17)(9.302,-4.000){2}{\rule{0.650pt}{0.400pt}}
\multiput(689.00,353.93)(1.378,-0.477){7}{\rule{1.140pt}{0.115pt}}
\multiput(689.00,354.17)(10.634,-5.000){2}{\rule{0.570pt}{0.400pt}}
\multiput(702.00,348.94)(1.797,-0.468){5}{\rule{1.400pt}{0.113pt}}
\multiput(702.00,349.17)(10.094,-4.000){2}{\rule{0.700pt}{0.400pt}}
\multiput(715.00,344.94)(1.651,-0.468){5}{\rule{1.300pt}{0.113pt}}
\multiput(715.00,345.17)(9.302,-4.000){2}{\rule{0.650pt}{0.400pt}}
\multiput(727.00,340.93)(1.378,-0.477){7}{\rule{1.140pt}{0.115pt}}
\multiput(727.00,341.17)(10.634,-5.000){2}{\rule{0.570pt}{0.400pt}}
\multiput(740.00,335.94)(1.797,-0.468){5}{\rule{1.400pt}{0.113pt}}
\multiput(740.00,336.17)(10.094,-4.000){2}{\rule{0.700pt}{0.400pt}}
\multiput(753.00,331.95)(2.695,-0.447){3}{\rule{1.833pt}{0.108pt}}
\multiput(753.00,332.17)(9.195,-3.000){2}{\rule{0.917pt}{0.400pt}}
\multiput(766.00,328.94)(1.651,-0.468){5}{\rule{1.300pt}{0.113pt}}
\multiput(766.00,329.17)(9.302,-4.000){2}{\rule{0.650pt}{0.400pt}}
\multiput(778.00,324.94)(1.797,-0.468){5}{\rule{1.400pt}{0.113pt}}
\multiput(778.00,325.17)(10.094,-4.000){2}{\rule{0.700pt}{0.400pt}}
\multiput(791.00,320.95)(2.695,-0.447){3}{\rule{1.833pt}{0.108pt}}
\multiput(791.00,321.17)(9.195,-3.000){2}{\rule{0.917pt}{0.400pt}}
\multiput(804.00,317.94)(1.651,-0.468){5}{\rule{1.300pt}{0.113pt}}
\multiput(804.00,318.17)(9.302,-4.000){2}{\rule{0.650pt}{0.400pt}}
\multiput(816.00,313.95)(2.695,-0.447){3}{\rule{1.833pt}{0.108pt}}
\multiput(816.00,314.17)(9.195,-3.000){2}{\rule{0.917pt}{0.400pt}}
\multiput(829.00,310.95)(2.695,-0.447){3}{\rule{1.833pt}{0.108pt}}
\multiput(829.00,311.17)(9.195,-3.000){2}{\rule{0.917pt}{0.400pt}}
\multiput(842.00,307.94)(1.651,-0.468){5}{\rule{1.300pt}{0.113pt}}
\multiput(842.00,308.17)(9.302,-4.000){2}{\rule{0.650pt}{0.400pt}}
\multiput(854.00,303.95)(2.695,-0.447){3}{\rule{1.833pt}{0.108pt}}
\multiput(854.00,304.17)(9.195,-3.000){2}{\rule{0.917pt}{0.400pt}}
\multiput(867.00,300.95)(2.695,-0.447){3}{\rule{1.833pt}{0.108pt}}
\multiput(867.00,301.17)(9.195,-3.000){2}{\rule{0.917pt}{0.400pt}}
\put(880,297.17){\rule{2.700pt}{0.400pt}}
\multiput(880.00,298.17)(7.396,-2.000){2}{\rule{1.350pt}{0.400pt}}
\multiput(893.00,295.95)(2.472,-0.447){3}{\rule{1.700pt}{0.108pt}}
\multiput(893.00,296.17)(8.472,-3.000){2}{\rule{0.850pt}{0.400pt}}
\multiput(905.00,292.95)(2.695,-0.447){3}{\rule{1.833pt}{0.108pt}}
\multiput(905.00,293.17)(9.195,-3.000){2}{\rule{0.917pt}{0.400pt}}
\put(918,289.17){\rule{2.700pt}{0.400pt}}
\multiput(918.00,290.17)(7.396,-2.000){2}{\rule{1.350pt}{0.400pt}}
\multiput(931.00,287.95)(2.472,-0.447){3}{\rule{1.700pt}{0.108pt}}
\multiput(931.00,288.17)(8.472,-3.000){2}{\rule{0.850pt}{0.400pt}}
\put(943,284.17){\rule{2.700pt}{0.400pt}}
\multiput(943.00,285.17)(7.396,-2.000){2}{\rule{1.350pt}{0.400pt}}
\put(956,282.17){\rule{2.700pt}{0.400pt}}
\multiput(956.00,283.17)(7.396,-2.000){2}{\rule{1.350pt}{0.400pt}}
\put(969,280.17){\rule{2.700pt}{0.400pt}}
\multiput(969.00,281.17)(7.396,-2.000){2}{\rule{1.350pt}{0.400pt}}
\multiput(982.00,278.95)(2.472,-0.447){3}{\rule{1.700pt}{0.108pt}}
\multiput(982.00,279.17)(8.472,-3.000){2}{\rule{0.850pt}{0.400pt}}
\put(994,275.17){\rule{2.700pt}{0.400pt}}
\multiput(994.00,276.17)(7.396,-2.000){2}{\rule{1.350pt}{0.400pt}}
\put(1007,273.67){\rule{3.132pt}{0.400pt}}
\multiput(1007.00,274.17)(6.500,-1.000){2}{\rule{1.566pt}{0.400pt}}
\put(1020,272.17){\rule{2.500pt}{0.400pt}}
\multiput(1020.00,273.17)(6.811,-2.000){2}{\rule{1.250pt}{0.400pt}}
\put(1032,270.17){\rule{2.700pt}{0.400pt}}
\multiput(1032.00,271.17)(7.396,-2.000){2}{\rule{1.350pt}{0.400pt}}
\put(1045,268.17){\rule{2.700pt}{0.400pt}}
\multiput(1045.00,269.17)(7.396,-2.000){2}{\rule{1.350pt}{0.400pt}}
\put(1058,266.67){\rule{2.891pt}{0.400pt}}
\multiput(1058.00,267.17)(6.000,-1.000){2}{\rule{1.445pt}{0.400pt}}
\put(1070,265.17){\rule{2.700pt}{0.400pt}}
\multiput(1070.00,266.17)(7.396,-2.000){2}{\rule{1.350pt}{0.400pt}}
\put(1083,263.67){\rule{3.132pt}{0.400pt}}
\multiput(1083.00,264.17)(6.500,-1.000){2}{\rule{1.566pt}{0.400pt}}
\put(1096,262.17){\rule{2.700pt}{0.400pt}}
\multiput(1096.00,263.17)(7.396,-2.000){2}{\rule{1.350pt}{0.400pt}}
\put(1109,260.67){\rule{2.891pt}{0.400pt}}
\multiput(1109.00,261.17)(6.000,-1.000){2}{\rule{1.445pt}{0.400pt}}
\put(1121,259.67){\rule{3.132pt}{0.400pt}}
\multiput(1121.00,260.17)(6.500,-1.000){2}{\rule{1.566pt}{0.400pt}}
\put(1134,258.17){\rule{2.700pt}{0.400pt}}
\multiput(1134.00,259.17)(7.396,-2.000){2}{\rule{1.350pt}{0.400pt}}
\put(1147,256.67){\rule{2.891pt}{0.400pt}}
\multiput(1147.00,257.17)(6.000,-1.000){2}{\rule{1.445pt}{0.400pt}}
\put(1159,255.67){\rule{3.132pt}{0.400pt}}
\multiput(1159.00,256.17)(6.500,-1.000){2}{\rule{1.566pt}{0.400pt}}
\put(1172,254.67){\rule{3.132pt}{0.400pt}}
\multiput(1172.00,255.17)(6.500,-1.000){2}{\rule{1.566pt}{0.400pt}}
\put(1185,253.67){\rule{3.132pt}{0.400pt}}
\multiput(1185.00,254.17)(6.500,-1.000){2}{\rule{1.566pt}{0.400pt}}
\put(1198,252.67){\rule{2.891pt}{0.400pt}}
\multiput(1198.00,253.17)(6.000,-1.000){2}{\rule{1.445pt}{0.400pt}}
\put(1210,251.67){\rule{3.132pt}{0.400pt}}
\multiput(1210.00,252.17)(6.500,-1.000){2}{\rule{1.566pt}{0.400pt}}
\put(1223,250.67){\rule{3.132pt}{0.400pt}}
\multiput(1223.00,251.17)(6.500,-1.000){2}{\rule{1.566pt}{0.400pt}}
\put(1248,249.67){\rule{3.132pt}{0.400pt}}
\multiput(1248.00,250.17)(6.500,-1.000){2}{\rule{1.566pt}{0.400pt}}
\put(1261,248.67){\rule{3.132pt}{0.400pt}}
\multiput(1261.00,249.17)(6.500,-1.000){2}{\rule{1.566pt}{0.400pt}}
\put(1274,247.67){\rule{3.132pt}{0.400pt}}
\multiput(1274.00,248.17)(6.500,-1.000){2}{\rule{1.566pt}{0.400pt}}
\put(1287,246.67){\rule{2.891pt}{0.400pt}}
\multiput(1287.00,247.17)(6.000,-1.000){2}{\rule{1.445pt}{0.400pt}}
\put(1236.0,251.0){\rule[-0.200pt]{2.891pt}{0.400pt}}
\put(1312,245.67){\rule{3.132pt}{0.400pt}}
\multiput(1312.00,246.17)(6.500,-1.000){2}{\rule{1.566pt}{0.400pt}}
\put(1325,244.67){\rule{2.891pt}{0.400pt}}
\multiput(1325.00,245.17)(6.000,-1.000){2}{\rule{1.445pt}{0.400pt}}
\put(1299.0,247.0){\rule[-0.200pt]{3.132pt}{0.400pt}}
\put(1350,243.67){\rule{3.132pt}{0.400pt}}
\multiput(1350.00,244.17)(6.500,-1.000){2}{\rule{1.566pt}{0.400pt}}
\put(1363,242.67){\rule{2.891pt}{0.400pt}}
\multiput(1363.00,243.17)(6.000,-1.000){2}{\rule{1.445pt}{0.400pt}}
\put(1337.0,245.0){\rule[-0.200pt]{3.132pt}{0.400pt}}
\put(1388,241.67){\rule{3.132pt}{0.400pt}}
\multiput(1388.00,242.17)(6.500,-1.000){2}{\rule{1.566pt}{0.400pt}}
\put(1401,240.67){\rule{3.132pt}{0.400pt}}
\multiput(1401.00,241.17)(6.500,-1.000){2}{\rule{1.566pt}{0.400pt}}
\put(1414,239.67){\rule{2.891pt}{0.400pt}}
\multiput(1414.00,240.17)(6.000,-1.000){2}{\rule{1.445pt}{0.400pt}}
\put(1375.0,243.0){\rule[-0.200pt]{3.132pt}{0.400pt}}
\put(1426.0,240.0){\rule[-0.200pt]{3.132pt}{0.400pt}}
\put(1279,779){\makebox(0,0)[r]{d=3}}
\multiput(1299,779)(20.756,0.000){5}{\usebox{\plotpoint}}
\put(1399,779){\usebox{\plotpoint}}
\put(181,639){\usebox{\plotpoint}}
\put(181.00,639.00){\usebox{\plotpoint}}
\put(197.96,627.03){\usebox{\plotpoint}}
\put(214.80,614.91){\usebox{\plotpoint}}
\put(231.86,603.10){\usebox{\plotpoint}}
\put(249.26,591.81){\usebox{\plotpoint}}
\put(266.11,579.69){\usebox{\plotpoint}}
\put(283.61,568.54){\usebox{\plotpoint}}
\put(300.55,556.59){\usebox{\plotpoint}}
\put(317.87,545.17){\usebox{\plotpoint}}
\put(335.15,533.68){\usebox{\plotpoint}}
\put(352.59,522.44){\usebox{\plotpoint}}
\put(370.04,511.21){\usebox{\plotpoint}}
\put(387.43,499.89){\usebox{\plotpoint}}
\put(405.11,489.01){\usebox{\plotpoint}}
\put(422.50,477.69){\usebox{\plotpoint}}
\put(440.36,467.12){\usebox{\plotpoint}}
\put(458.28,456.67){\usebox{\plotpoint}}
\put(475.77,445.51){\usebox{\plotpoint}}
\put(494.04,435.67){\usebox{\plotpoint}}
\put(511.62,424.67){\usebox{\plotpoint}}
\put(529.89,414.83){\usebox{\plotpoint}}
\put(548.17,404.99){\usebox{\plotpoint}}
\put(566.34,395.00){\usebox{\plotpoint}}
\put(584.88,385.68){\usebox{\plotpoint}}
\put(603.01,375.61){\usebox{\plotpoint}}
\put(621.59,366.38){\usebox{\plotpoint}}
\put(640.11,357.02){\usebox{\plotpoint}}
\put(658.96,348.33){\usebox{\plotpoint}}
\put(677.79,339.60){\usebox{\plotpoint}}
\put(696.47,330.55){\usebox{\plotpoint}}
\put(715.66,322.67){\usebox{\plotpoint}}
\put(734.54,314.10){\usebox{\plotpoint}}
\put(753.88,306.59){\usebox{\plotpoint}}
\put(772.84,298.15){\usebox{\plotpoint}}
\put(792.18,290.64){\usebox{\plotpoint}}
\put(811.75,283.77){\usebox{\plotpoint}}
\put(831.12,276.35){\usebox{\plotpoint}}
\put(850.89,270.04){\usebox{\plotpoint}}
\put(870.39,262.96){\usebox{\plotpoint}}
\put(890.23,256.85){\usebox{\plotpoint}}
\put(909.98,250.47){\usebox{\plotpoint}}
\put(930.05,245.22){\usebox{\plotpoint}}
\put(949.95,239.40){\usebox{\plotpoint}}
\put(969.92,233.79){\usebox{\plotpoint}}
\put(990.11,228.97){\usebox{\plotpoint}}
\put(1010.31,224.24){\usebox{\plotpoint}}
\put(1030.49,219.38){\usebox{\plotpoint}}
\put(1050.89,215.64){\usebox{\plotpoint}}
\put(1071.26,211.71){\usebox{\plotpoint}}
\put(1091.61,207.68){\usebox{\plotpoint}}
\put(1112.11,204.48){\usebox{\plotpoint}}
\put(1132.61,201.21){\usebox{\plotpoint}}
\put(1153.11,197.98){\usebox{\plotpoint}}
\put(1173.73,195.73){\usebox{\plotpoint}}
\put(1194.32,193.28){\usebox{\plotpoint}}
\put(1214.89,190.62){\usebox{\plotpoint}}
\put(1235.58,189.03){\usebox{\plotpoint}}
\put(1256.27,187.36){\usebox{\plotpoint}}
\put(1276.97,185.77){\usebox{\plotpoint}}
\put(1297.65,184.11){\usebox{\plotpoint}}
\put(1318.39,183.51){\usebox{\plotpoint}}
\put(1339.12,182.84){\usebox{\plotpoint}}
\put(1359.84,182.00){\usebox{\plotpoint}}
\put(1380.55,181.00){\usebox{\plotpoint}}
\put(1401.31,181.00){\usebox{\plotpoint}}
\put(1422.06,181.00){\usebox{\plotpoint}}
\put(1439,181){\usebox{\plotpoint}}
\sbox{\plotpoint}{\rule[-0.400pt]{0.800pt}{0.800pt}}%
\put(1279,738){\makebox(0,0)[r]{d=4}}
\put(1299.0,738.0){\rule[-0.400pt]{24.090pt}{0.800pt}}
\put(181,651){\usebox{\plotpoint}}
\multiput(181.00,649.07)(1.244,-0.536){5}{\rule{1.933pt}{0.129pt}}
\multiput(181.00,649.34)(8.987,-6.000){2}{\rule{0.967pt}{0.800pt}}
\multiput(194.00,643.08)(0.913,-0.526){7}{\rule{1.571pt}{0.127pt}}
\multiput(194.00,643.34)(8.738,-7.000){2}{\rule{0.786pt}{0.800pt}}
\multiput(206.00,636.08)(1.000,-0.526){7}{\rule{1.686pt}{0.127pt}}
\multiput(206.00,636.34)(9.501,-7.000){2}{\rule{0.843pt}{0.800pt}}
\multiput(219.00,629.07)(1.244,-0.536){5}{\rule{1.933pt}{0.129pt}}
\multiput(219.00,629.34)(8.987,-6.000){2}{\rule{0.967pt}{0.800pt}}
\multiput(232.00,623.08)(1.000,-0.526){7}{\rule{1.686pt}{0.127pt}}
\multiput(232.00,623.34)(9.501,-7.000){2}{\rule{0.843pt}{0.800pt}}
\multiput(245.00,616.08)(0.913,-0.526){7}{\rule{1.571pt}{0.127pt}}
\multiput(245.00,616.34)(8.738,-7.000){2}{\rule{0.786pt}{0.800pt}}
\multiput(257.00,609.08)(1.000,-0.526){7}{\rule{1.686pt}{0.127pt}}
\multiput(257.00,609.34)(9.501,-7.000){2}{\rule{0.843pt}{0.800pt}}
\multiput(270.00,602.08)(1.000,-0.526){7}{\rule{1.686pt}{0.127pt}}
\multiput(270.00,602.34)(9.501,-7.000){2}{\rule{0.843pt}{0.800pt}}
\multiput(283.00,595.08)(0.774,-0.520){9}{\rule{1.400pt}{0.125pt}}
\multiput(283.00,595.34)(9.094,-8.000){2}{\rule{0.700pt}{0.800pt}}
\multiput(295.00,587.08)(1.000,-0.526){7}{\rule{1.686pt}{0.127pt}}
\multiput(295.00,587.34)(9.501,-7.000){2}{\rule{0.843pt}{0.800pt}}
\multiput(308.00,580.08)(1.000,-0.526){7}{\rule{1.686pt}{0.127pt}}
\multiput(308.00,580.34)(9.501,-7.000){2}{\rule{0.843pt}{0.800pt}}
\multiput(321.00,573.08)(0.913,-0.526){7}{\rule{1.571pt}{0.127pt}}
\multiput(321.00,573.34)(8.738,-7.000){2}{\rule{0.786pt}{0.800pt}}
\multiput(333.00,566.08)(0.847,-0.520){9}{\rule{1.500pt}{0.125pt}}
\multiput(333.00,566.34)(9.887,-8.000){2}{\rule{0.750pt}{0.800pt}}
\multiput(346.00,558.08)(1.000,-0.526){7}{\rule{1.686pt}{0.127pt}}
\multiput(346.00,558.34)(9.501,-7.000){2}{\rule{0.843pt}{0.800pt}}
\multiput(359.00,551.08)(0.847,-0.520){9}{\rule{1.500pt}{0.125pt}}
\multiput(359.00,551.34)(9.887,-8.000){2}{\rule{0.750pt}{0.800pt}}
\multiput(372.00,543.08)(0.913,-0.526){7}{\rule{1.571pt}{0.127pt}}
\multiput(372.00,543.34)(8.738,-7.000){2}{\rule{0.786pt}{0.800pt}}
\multiput(384.00,536.08)(1.000,-0.526){7}{\rule{1.686pt}{0.127pt}}
\multiput(384.00,536.34)(9.501,-7.000){2}{\rule{0.843pt}{0.800pt}}
\multiput(397.00,529.08)(0.847,-0.520){9}{\rule{1.500pt}{0.125pt}}
\multiput(397.00,529.34)(9.887,-8.000){2}{\rule{0.750pt}{0.800pt}}
\multiput(410.00,521.08)(0.913,-0.526){7}{\rule{1.571pt}{0.127pt}}
\multiput(410.00,521.34)(8.738,-7.000){2}{\rule{0.786pt}{0.800pt}}
\multiput(422.00,514.08)(0.847,-0.520){9}{\rule{1.500pt}{0.125pt}}
\multiput(422.00,514.34)(9.887,-8.000){2}{\rule{0.750pt}{0.800pt}}
\multiput(435.00,506.08)(1.000,-0.526){7}{\rule{1.686pt}{0.127pt}}
\multiput(435.00,506.34)(9.501,-7.000){2}{\rule{0.843pt}{0.800pt}}
\multiput(448.00,499.08)(0.847,-0.520){9}{\rule{1.500pt}{0.125pt}}
\multiput(448.00,499.34)(9.887,-8.000){2}{\rule{0.750pt}{0.800pt}}
\multiput(461.00,491.08)(0.913,-0.526){7}{\rule{1.571pt}{0.127pt}}
\multiput(461.00,491.34)(8.738,-7.000){2}{\rule{0.786pt}{0.800pt}}
\multiput(473.00,484.08)(1.000,-0.526){7}{\rule{1.686pt}{0.127pt}}
\multiput(473.00,484.34)(9.501,-7.000){2}{\rule{0.843pt}{0.800pt}}
\multiput(486.00,477.08)(0.847,-0.520){9}{\rule{1.500pt}{0.125pt}}
\multiput(486.00,477.34)(9.887,-8.000){2}{\rule{0.750pt}{0.800pt}}
\multiput(499.00,469.08)(0.913,-0.526){7}{\rule{1.571pt}{0.127pt}}
\multiput(499.00,469.34)(8.738,-7.000){2}{\rule{0.786pt}{0.800pt}}
\multiput(511.00,462.08)(0.847,-0.520){9}{\rule{1.500pt}{0.125pt}}
\multiput(511.00,462.34)(9.887,-8.000){2}{\rule{0.750pt}{0.800pt}}
\multiput(524.00,454.08)(1.000,-0.526){7}{\rule{1.686pt}{0.127pt}}
\multiput(524.00,454.34)(9.501,-7.000){2}{\rule{0.843pt}{0.800pt}}
\multiput(537.00,447.08)(1.000,-0.526){7}{\rule{1.686pt}{0.127pt}}
\multiput(537.00,447.34)(9.501,-7.000){2}{\rule{0.843pt}{0.800pt}}
\multiput(550.00,440.08)(0.913,-0.526){7}{\rule{1.571pt}{0.127pt}}
\multiput(550.00,440.34)(8.738,-7.000){2}{\rule{0.786pt}{0.800pt}}
\multiput(562.00,433.08)(1.000,-0.526){7}{\rule{1.686pt}{0.127pt}}
\multiput(562.00,433.34)(9.501,-7.000){2}{\rule{0.843pt}{0.800pt}}
\multiput(575.00,426.08)(1.000,-0.526){7}{\rule{1.686pt}{0.127pt}}
\multiput(575.00,426.34)(9.501,-7.000){2}{\rule{0.843pt}{0.800pt}}
\multiput(588.00,419.08)(0.913,-0.526){7}{\rule{1.571pt}{0.127pt}}
\multiput(588.00,419.34)(8.738,-7.000){2}{\rule{0.786pt}{0.800pt}}
\multiput(600.00,412.08)(1.000,-0.526){7}{\rule{1.686pt}{0.127pt}}
\multiput(600.00,412.34)(9.501,-7.000){2}{\rule{0.843pt}{0.800pt}}
\multiput(613.00,405.08)(1.000,-0.526){7}{\rule{1.686pt}{0.127pt}}
\multiput(613.00,405.34)(9.501,-7.000){2}{\rule{0.843pt}{0.800pt}}
\multiput(626.00,398.08)(0.913,-0.526){7}{\rule{1.571pt}{0.127pt}}
\multiput(626.00,398.34)(8.738,-7.000){2}{\rule{0.786pt}{0.800pt}}
\multiput(638.00,391.08)(1.000,-0.526){7}{\rule{1.686pt}{0.127pt}}
\multiput(638.00,391.34)(9.501,-7.000){2}{\rule{0.843pt}{0.800pt}}
\multiput(651.00,384.08)(1.000,-0.526){7}{\rule{1.686pt}{0.127pt}}
\multiput(651.00,384.34)(9.501,-7.000){2}{\rule{0.843pt}{0.800pt}}
\multiput(664.00,377.07)(1.244,-0.536){5}{\rule{1.933pt}{0.129pt}}
\multiput(664.00,377.34)(8.987,-6.000){2}{\rule{0.967pt}{0.800pt}}
\multiput(677.00,371.08)(0.913,-0.526){7}{\rule{1.571pt}{0.127pt}}
\multiput(677.00,371.34)(8.738,-7.000){2}{\rule{0.786pt}{0.800pt}}
\multiput(689.00,364.07)(1.244,-0.536){5}{\rule{1.933pt}{0.129pt}}
\multiput(689.00,364.34)(8.987,-6.000){2}{\rule{0.967pt}{0.800pt}}
\multiput(702.00,358.08)(1.000,-0.526){7}{\rule{1.686pt}{0.127pt}}
\multiput(702.00,358.34)(9.501,-7.000){2}{\rule{0.843pt}{0.800pt}}
\multiput(715.00,351.07)(1.132,-0.536){5}{\rule{1.800pt}{0.129pt}}
\multiput(715.00,351.34)(8.264,-6.000){2}{\rule{0.900pt}{0.800pt}}
\multiput(727.00,345.07)(1.244,-0.536){5}{\rule{1.933pt}{0.129pt}}
\multiput(727.00,345.34)(8.987,-6.000){2}{\rule{0.967pt}{0.800pt}}
\multiput(740.00,339.07)(1.244,-0.536){5}{\rule{1.933pt}{0.129pt}}
\multiput(740.00,339.34)(8.987,-6.000){2}{\rule{0.967pt}{0.800pt}}
\multiput(753.00,333.07)(1.244,-0.536){5}{\rule{1.933pt}{0.129pt}}
\multiput(753.00,333.34)(8.987,-6.000){2}{\rule{0.967pt}{0.800pt}}
\multiput(766.00,327.07)(1.132,-0.536){5}{\rule{1.800pt}{0.129pt}}
\multiput(766.00,327.34)(8.264,-6.000){2}{\rule{0.900pt}{0.800pt}}
\multiput(778.00,321.07)(1.244,-0.536){5}{\rule{1.933pt}{0.129pt}}
\multiput(778.00,321.34)(8.987,-6.000){2}{\rule{0.967pt}{0.800pt}}
\multiput(791.00,315.07)(1.244,-0.536){5}{\rule{1.933pt}{0.129pt}}
\multiput(791.00,315.34)(8.987,-6.000){2}{\rule{0.967pt}{0.800pt}}
\multiput(804.00,309.06)(1.600,-0.560){3}{\rule{2.120pt}{0.135pt}}
\multiput(804.00,309.34)(7.600,-5.000){2}{\rule{1.060pt}{0.800pt}}
\multiput(816.00,304.07)(1.244,-0.536){5}{\rule{1.933pt}{0.129pt}}
\multiput(816.00,304.34)(8.987,-6.000){2}{\rule{0.967pt}{0.800pt}}
\multiput(829.00,298.06)(1.768,-0.560){3}{\rule{2.280pt}{0.135pt}}
\multiput(829.00,298.34)(8.268,-5.000){2}{\rule{1.140pt}{0.800pt}}
\multiput(842.00,293.06)(1.600,-0.560){3}{\rule{2.120pt}{0.135pt}}
\multiput(842.00,293.34)(7.600,-5.000){2}{\rule{1.060pt}{0.800pt}}
\multiput(854.00,288.06)(1.768,-0.560){3}{\rule{2.280pt}{0.135pt}}
\multiput(854.00,288.34)(8.268,-5.000){2}{\rule{1.140pt}{0.800pt}}
\multiput(867.00,283.07)(1.244,-0.536){5}{\rule{1.933pt}{0.129pt}}
\multiput(867.00,283.34)(8.987,-6.000){2}{\rule{0.967pt}{0.800pt}}
\put(880,275.34){\rule{2.800pt}{0.800pt}}
\multiput(880.00,277.34)(7.188,-4.000){2}{\rule{1.400pt}{0.800pt}}
\multiput(893.00,273.06)(1.600,-0.560){3}{\rule{2.120pt}{0.135pt}}
\multiput(893.00,273.34)(7.600,-5.000){2}{\rule{1.060pt}{0.800pt}}
\multiput(905.00,268.06)(1.768,-0.560){3}{\rule{2.280pt}{0.135pt}}
\multiput(905.00,268.34)(8.268,-5.000){2}{\rule{1.140pt}{0.800pt}}
\multiput(918.00,263.06)(1.768,-0.560){3}{\rule{2.280pt}{0.135pt}}
\multiput(918.00,263.34)(8.268,-5.000){2}{\rule{1.140pt}{0.800pt}}
\put(931,256.34){\rule{2.600pt}{0.800pt}}
\multiput(931.00,258.34)(6.604,-4.000){2}{\rule{1.300pt}{0.800pt}}
\put(943,252.34){\rule{2.800pt}{0.800pt}}
\multiput(943.00,254.34)(7.188,-4.000){2}{\rule{1.400pt}{0.800pt}}
\multiput(956.00,250.06)(1.768,-0.560){3}{\rule{2.280pt}{0.135pt}}
\multiput(956.00,250.34)(8.268,-5.000){2}{\rule{1.140pt}{0.800pt}}
\put(969,243.34){\rule{2.800pt}{0.800pt}}
\multiput(969.00,245.34)(7.188,-4.000){2}{\rule{1.400pt}{0.800pt}}
\put(982,239.34){\rule{2.600pt}{0.800pt}}
\multiput(982.00,241.34)(6.604,-4.000){2}{\rule{1.300pt}{0.800pt}}
\put(994,235.34){\rule{2.800pt}{0.800pt}}
\multiput(994.00,237.34)(7.188,-4.000){2}{\rule{1.400pt}{0.800pt}}
\put(1007,231.84){\rule{3.132pt}{0.800pt}}
\multiput(1007.00,233.34)(6.500,-3.000){2}{\rule{1.566pt}{0.800pt}}
\put(1020,228.34){\rule{2.600pt}{0.800pt}}
\multiput(1020.00,230.34)(6.604,-4.000){2}{\rule{1.300pt}{0.800pt}}
\put(1032,224.34){\rule{2.800pt}{0.800pt}}
\multiput(1032.00,226.34)(7.188,-4.000){2}{\rule{1.400pt}{0.800pt}}
\put(1045,220.84){\rule{3.132pt}{0.800pt}}
\multiput(1045.00,222.34)(6.500,-3.000){2}{\rule{1.566pt}{0.800pt}}
\put(1058,217.84){\rule{2.891pt}{0.800pt}}
\multiput(1058.00,219.34)(6.000,-3.000){2}{\rule{1.445pt}{0.800pt}}
\put(1070,214.84){\rule{3.132pt}{0.800pt}}
\multiput(1070.00,216.34)(6.500,-3.000){2}{\rule{1.566pt}{0.800pt}}
\put(1083,211.34){\rule{2.800pt}{0.800pt}}
\multiput(1083.00,213.34)(7.188,-4.000){2}{\rule{1.400pt}{0.800pt}}
\put(1096,208.34){\rule{3.132pt}{0.800pt}}
\multiput(1096.00,209.34)(6.500,-2.000){2}{\rule{1.566pt}{0.800pt}}
\put(1109,205.84){\rule{2.891pt}{0.800pt}}
\multiput(1109.00,207.34)(6.000,-3.000){2}{\rule{1.445pt}{0.800pt}}
\put(1121,202.84){\rule{3.132pt}{0.800pt}}
\multiput(1121.00,204.34)(6.500,-3.000){2}{\rule{1.566pt}{0.800pt}}
\put(1134,199.84){\rule{3.132pt}{0.800pt}}
\multiput(1134.00,201.34)(6.500,-3.000){2}{\rule{1.566pt}{0.800pt}}
\put(1147,197.34){\rule{2.891pt}{0.800pt}}
\multiput(1147.00,198.34)(6.000,-2.000){2}{\rule{1.445pt}{0.800pt}}
\put(1159,195.34){\rule{3.132pt}{0.800pt}}
\multiput(1159.00,196.34)(6.500,-2.000){2}{\rule{1.566pt}{0.800pt}}
\put(1172,192.84){\rule{3.132pt}{0.800pt}}
\multiput(1172.00,194.34)(6.500,-3.000){2}{\rule{1.566pt}{0.800pt}}
\put(1185,190.34){\rule{3.132pt}{0.800pt}}
\multiput(1185.00,191.34)(6.500,-2.000){2}{\rule{1.566pt}{0.800pt}}
\put(1198,188.34){\rule{2.891pt}{0.800pt}}
\multiput(1198.00,189.34)(6.000,-2.000){2}{\rule{1.445pt}{0.800pt}}
\put(1210,186.34){\rule{3.132pt}{0.800pt}}
\multiput(1210.00,187.34)(6.500,-2.000){2}{\rule{1.566pt}{0.800pt}}
\put(1223,184.84){\rule{3.132pt}{0.800pt}}
\multiput(1223.00,185.34)(6.500,-1.000){2}{\rule{1.566pt}{0.800pt}}
\put(1236,183.34){\rule{2.891pt}{0.800pt}}
\multiput(1236.00,184.34)(6.000,-2.000){2}{\rule{1.445pt}{0.800pt}}
\put(1248,181.34){\rule{3.132pt}{0.800pt}}
\multiput(1248.00,182.34)(6.500,-2.000){2}{\rule{1.566pt}{0.800pt}}
\put(1261,179.84){\rule{3.132pt}{0.800pt}}
\multiput(1261.00,180.34)(6.500,-1.000){2}{\rule{1.566pt}{0.800pt}}
\put(1274,178.34){\rule{3.132pt}{0.800pt}}
\multiput(1274.00,179.34)(6.500,-2.000){2}{\rule{1.566pt}{0.800pt}}
\put(1287,176.84){\rule{2.891pt}{0.800pt}}
\multiput(1287.00,177.34)(6.000,-1.000){2}{\rule{1.445pt}{0.800pt}}
\put(1299,175.84){\rule{3.132pt}{0.800pt}}
\multiput(1299.00,176.34)(6.500,-1.000){2}{\rule{1.566pt}{0.800pt}}
\put(1312,174.84){\rule{3.132pt}{0.800pt}}
\multiput(1312.00,175.34)(6.500,-1.000){2}{\rule{1.566pt}{0.800pt}}
\put(1325,173.84){\rule{2.891pt}{0.800pt}}
\multiput(1325.00,174.34)(6.000,-1.000){2}{\rule{1.445pt}{0.800pt}}
\put(1337,172.84){\rule{3.132pt}{0.800pt}}
\multiput(1337.00,173.34)(6.500,-1.000){2}{\rule{1.566pt}{0.800pt}}
\put(1350,171.84){\rule{3.132pt}{0.800pt}}
\multiput(1350.00,172.34)(6.500,-1.000){2}{\rule{1.566pt}{0.800pt}}
\put(1363,170.84){\rule{2.891pt}{0.800pt}}
\multiput(1363.00,171.34)(6.000,-1.000){2}{\rule{1.445pt}{0.800pt}}
\put(1388,169.84){\rule{3.132pt}{0.800pt}}
\multiput(1388.00,170.34)(6.500,-1.000){2}{\rule{1.566pt}{0.800pt}}
\put(1375.0,172.0){\rule[-0.400pt]{3.132pt}{0.800pt}}
\put(1414,168.84){\rule{2.891pt}{0.800pt}}
\multiput(1414.00,169.34)(6.000,-1.000){2}{\rule{1.445pt}{0.800pt}}
\put(1401.0,171.0){\rule[-0.400pt]{3.132pt}{0.800pt}}
\put(1426.0,170.0){\rule[-0.400pt]{3.132pt}{0.800pt}}
\sbox{\plotpoint}{\rule[-0.500pt]{1.000pt}{1.000pt}}%
\put(1279,697){\makebox(0,0)[r]{d=5}}
\multiput(1299,697)(20.756,0.000){5}{\usebox{\plotpoint}}
\put(1399,697){\usebox{\plotpoint}}
\put(181,651){\usebox{\plotpoint}}
\put(181.00,651.00){\usebox{\plotpoint}}
\put(200.60,644.25){\usebox{\plotpoint}}
\put(219.92,636.65){\usebox{\plotpoint}}
\put(239.29,629.20){\usebox{\plotpoint}}
\put(258.14,620.56){\usebox{\plotpoint}}
\put(277.31,612.63){\usebox{\plotpoint}}
\put(295.97,603.55){\usebox{\plotpoint}}
\put(314.81,594.85){\usebox{\plotpoint}}
\put(333.48,585.78){\usebox{\plotpoint}}
\put(352.13,576.70){\usebox{\plotpoint}}
\put(370.76,567.57){\usebox{\plotpoint}}
\put(388.99,557.70){\usebox{\plotpoint}}
\put(407.51,548.34){\usebox{\plotpoint}}
\put(425.55,538.09){\usebox{\plotpoint}}
\put(443.83,528.25){\usebox{\plotpoint}}
\put(462.12,518.44){\usebox{\plotpoint}}
\put(480.56,508.93){\usebox{\plotpoint}}
\put(498.84,499.09){\usebox{\plotpoint}}
\put(516.69,488.50){\usebox{\plotpoint}}
\put(534.72,478.23){\usebox{\plotpoint}}
\put(552.93,468.29){\usebox{\plotpoint}}
\put(571.03,458.14){\usebox{\plotpoint}}
\put(589.28,448.25){\usebox{\plotpoint}}
\put(607.35,438.04){\usebox{\plotpoint}}
\put(625.63,428.20){\usebox{\plotpoint}}
\put(643.67,417.95){\usebox{\plotpoint}}
\put(662.28,408.79){\usebox{\plotpoint}}
\put(680.54,398.93){\usebox{\plotpoint}}
\put(698.65,388.80){\usebox{\plotpoint}}
\put(717.28,379.67){\usebox{\plotpoint}}
\put(735.63,370.02){\usebox{\plotpoint}}
\put(754.07,360.51){\usebox{\plotpoint}}
\put(772.57,351.16){\usebox{\plotpoint}}
\put(791.14,341.93){\usebox{\plotpoint}}
\put(809.90,333.05){\usebox{\plotpoint}}
\put(828.65,324.16){\usebox{\plotpoint}}
\put(847.41,315.29){\usebox{\plotpoint}}
\put(866.16,306.39){\usebox{\plotpoint}}
\put(885.36,298.53){\usebox{\plotpoint}}
\put(904.39,290.25){\usebox{\plotpoint}}
\put(923.60,282.42){\usebox{\plotpoint}}
\put(942.63,274.15){\usebox{\plotpoint}}
\put(962.00,266.69){\usebox{\plotpoint}}
\put(981.67,260.10){\usebox{\plotpoint}}
\put(1000.92,252.34){\usebox{\plotpoint}}
\put(1020.60,245.80){\usebox{\plotpoint}}
\put(1040.16,238.86){\usebox{\plotpoint}}
\put(1059.87,232.38){\usebox{\plotpoint}}
\put(1079.82,226.73){\usebox{\plotpoint}}
\put(1099.72,220.86){\usebox{\plotpoint}}
\put(1119.71,215.32){\usebox{\plotpoint}}
\put(1139.82,210.21){\usebox{\plotpoint}}
\put(1159.85,204.80){\usebox{\plotpoint}}
\put(1180.07,200.14){\usebox{\plotpoint}}
\put(1200.47,196.38){\usebox{\plotpoint}}
\put(1220.81,192.34){\usebox{\plotpoint}}
\put(1241.12,188.15){\usebox{\plotpoint}}
\put(1261.62,184.90){\usebox{\plotpoint}}
\put(1282.14,181.75){\usebox{\plotpoint}}
\put(1302.66,178.72){\usebox{\plotpoint}}
\put(1323.26,176.27){\usebox{\plotpoint}}
\put(1343.87,173.94){\usebox{\plotpoint}}
\put(1364.51,171.87){\usebox{\plotpoint}}
\put(1385.20,170.22){\usebox{\plotpoint}}
\put(1405.91,169.00){\usebox{\plotpoint}}
\put(1426.62,167.95){\usebox{\plotpoint}}
\put(1439,167){\usebox{\plotpoint}}
\sbox{\plotpoint}{\rule[-0.600pt]{1.200pt}{1.200pt}}%
\put(1279,656){\makebox(0,0)[r]{d=6}}
\put(1299.0,656.0){\rule[-0.600pt]{24.090pt}{1.200pt}}
\put(181,663){\usebox{\plotpoint}}
\put(181,659.01){\rule{3.132pt}{1.200pt}}
\multiput(181.00,660.51)(6.500,-3.000){2}{\rule{1.566pt}{1.200pt}}
\put(194,655.51){\rule{2.891pt}{1.200pt}}
\multiput(194.00,657.51)(6.000,-4.000){2}{\rule{1.445pt}{1.200pt}}
\put(206,651.51){\rule{3.132pt}{1.200pt}}
\multiput(206.00,653.51)(6.500,-4.000){2}{\rule{1.566pt}{1.200pt}}
\put(219,647.51){\rule{3.132pt}{1.200pt}}
\multiput(219.00,649.51)(6.500,-4.000){2}{\rule{1.566pt}{1.200pt}}
\put(232,643.51){\rule{3.132pt}{1.200pt}}
\multiput(232.00,645.51)(6.500,-4.000){2}{\rule{1.566pt}{1.200pt}}
\put(245,639.01){\rule{2.891pt}{1.200pt}}
\multiput(245.00,641.51)(6.000,-5.000){2}{\rule{1.445pt}{1.200pt}}
\put(257,634.01){\rule{3.132pt}{1.200pt}}
\multiput(257.00,636.51)(6.500,-5.000){2}{\rule{1.566pt}{1.200pt}}
\put(270,629.51){\rule{3.132pt}{1.200pt}}
\multiput(270.00,631.51)(6.500,-4.000){2}{\rule{1.566pt}{1.200pt}}
\put(283,625.01){\rule{2.891pt}{1.200pt}}
\multiput(283.00,627.51)(6.000,-5.000){2}{\rule{1.445pt}{1.200pt}}
\multiput(295.00,622.25)(0.962,-0.509){2}{\rule{2.900pt}{0.123pt}}
\multiput(295.00,622.51)(6.981,-6.000){2}{\rule{1.450pt}{1.200pt}}
\put(308,614.01){\rule{3.132pt}{1.200pt}}
\multiput(308.00,616.51)(6.500,-5.000){2}{\rule{1.566pt}{1.200pt}}
\put(321,609.01){\rule{2.891pt}{1.200pt}}
\multiput(321.00,611.51)(6.000,-5.000){2}{\rule{1.445pt}{1.200pt}}
\multiput(333.00,606.25)(0.962,-0.509){2}{\rule{2.900pt}{0.123pt}}
\multiput(333.00,606.51)(6.981,-6.000){2}{\rule{1.450pt}{1.200pt}}
\multiput(346.00,600.25)(0.962,-0.509){2}{\rule{2.900pt}{0.123pt}}
\multiput(346.00,600.51)(6.981,-6.000){2}{\rule{1.450pt}{1.200pt}}
\put(359,592.01){\rule{3.132pt}{1.200pt}}
\multiput(359.00,594.51)(6.500,-5.000){2}{\rule{1.566pt}{1.200pt}}
\multiput(372.00,589.25)(0.792,-0.509){2}{\rule{2.700pt}{0.123pt}}
\multiput(372.00,589.51)(6.396,-6.000){2}{\rule{1.350pt}{1.200pt}}
\multiput(384.00,583.25)(0.962,-0.509){2}{\rule{2.900pt}{0.123pt}}
\multiput(384.00,583.51)(6.981,-6.000){2}{\rule{1.450pt}{1.200pt}}
\multiput(397.00,577.25)(0.962,-0.509){2}{\rule{2.900pt}{0.123pt}}
\multiput(397.00,577.51)(6.981,-6.000){2}{\rule{1.450pt}{1.200pt}}
\multiput(410.00,571.26)(0.738,-0.505){4}{\rule{2.357pt}{0.122pt}}
\multiput(410.00,571.51)(7.108,-7.000){2}{\rule{1.179pt}{1.200pt}}
\multiput(422.00,564.25)(0.962,-0.509){2}{\rule{2.900pt}{0.123pt}}
\multiput(422.00,564.51)(6.981,-6.000){2}{\rule{1.450pt}{1.200pt}}
\multiput(435.00,558.25)(0.962,-0.509){2}{\rule{2.900pt}{0.123pt}}
\multiput(435.00,558.51)(6.981,-6.000){2}{\rule{1.450pt}{1.200pt}}
\multiput(448.00,552.26)(0.835,-0.505){4}{\rule{2.529pt}{0.122pt}}
\multiput(448.00,552.51)(7.752,-7.000){2}{\rule{1.264pt}{1.200pt}}
\multiput(461.00,545.25)(0.792,-0.509){2}{\rule{2.700pt}{0.123pt}}
\multiput(461.00,545.51)(6.396,-6.000){2}{\rule{1.350pt}{1.200pt}}
\multiput(473.00,539.26)(0.835,-0.505){4}{\rule{2.529pt}{0.122pt}}
\multiput(473.00,539.51)(7.752,-7.000){2}{\rule{1.264pt}{1.200pt}}
\multiput(486.00,532.25)(0.962,-0.509){2}{\rule{2.900pt}{0.123pt}}
\multiput(486.00,532.51)(6.981,-6.000){2}{\rule{1.450pt}{1.200pt}}
\multiput(499.00,526.26)(0.738,-0.505){4}{\rule{2.357pt}{0.122pt}}
\multiput(499.00,526.51)(7.108,-7.000){2}{\rule{1.179pt}{1.200pt}}
\multiput(511.00,519.26)(0.835,-0.505){4}{\rule{2.529pt}{0.122pt}}
\multiput(511.00,519.51)(7.752,-7.000){2}{\rule{1.264pt}{1.200pt}}
\multiput(524.00,512.25)(0.962,-0.509){2}{\rule{2.900pt}{0.123pt}}
\multiput(524.00,512.51)(6.981,-6.000){2}{\rule{1.450pt}{1.200pt}}
\multiput(537.00,506.26)(0.835,-0.505){4}{\rule{2.529pt}{0.122pt}}
\multiput(537.00,506.51)(7.752,-7.000){2}{\rule{1.264pt}{1.200pt}}
\multiput(550.00,499.26)(0.738,-0.505){4}{\rule{2.357pt}{0.122pt}}
\multiput(550.00,499.51)(7.108,-7.000){2}{\rule{1.179pt}{1.200pt}}
\multiput(562.00,492.26)(0.835,-0.505){4}{\rule{2.529pt}{0.122pt}}
\multiput(562.00,492.51)(7.752,-7.000){2}{\rule{1.264pt}{1.200pt}}
\multiput(575.00,485.26)(0.835,-0.505){4}{\rule{2.529pt}{0.122pt}}
\multiput(575.00,485.51)(7.752,-7.000){2}{\rule{1.264pt}{1.200pt}}
\multiput(588.00,478.26)(0.738,-0.505){4}{\rule{2.357pt}{0.122pt}}
\multiput(588.00,478.51)(7.108,-7.000){2}{\rule{1.179pt}{1.200pt}}
\multiput(600.00,471.25)(0.962,-0.509){2}{\rule{2.900pt}{0.123pt}}
\multiput(600.00,471.51)(6.981,-6.000){2}{\rule{1.450pt}{1.200pt}}
\multiput(613.00,465.26)(0.835,-0.505){4}{\rule{2.529pt}{0.122pt}}
\multiput(613.00,465.51)(7.752,-7.000){2}{\rule{1.264pt}{1.200pt}}
\multiput(626.00,458.26)(0.738,-0.505){4}{\rule{2.357pt}{0.122pt}}
\multiput(626.00,458.51)(7.108,-7.000){2}{\rule{1.179pt}{1.200pt}}
\multiput(638.00,451.26)(0.835,-0.505){4}{\rule{2.529pt}{0.122pt}}
\multiput(638.00,451.51)(7.752,-7.000){2}{\rule{1.264pt}{1.200pt}}
\multiput(651.00,444.26)(0.835,-0.505){4}{\rule{2.529pt}{0.122pt}}
\multiput(651.00,444.51)(7.752,-7.000){2}{\rule{1.264pt}{1.200pt}}
\multiput(664.00,437.26)(0.835,-0.505){4}{\rule{2.529pt}{0.122pt}}
\multiput(664.00,437.51)(7.752,-7.000){2}{\rule{1.264pt}{1.200pt}}
\multiput(677.00,430.26)(0.738,-0.505){4}{\rule{2.357pt}{0.122pt}}
\multiput(677.00,430.51)(7.108,-7.000){2}{\rule{1.179pt}{1.200pt}}
\multiput(689.00,423.26)(0.835,-0.505){4}{\rule{2.529pt}{0.122pt}}
\multiput(689.00,423.51)(7.752,-7.000){2}{\rule{1.264pt}{1.200pt}}
\multiput(702.00,416.25)(0.962,-0.509){2}{\rule{2.900pt}{0.123pt}}
\multiput(702.00,416.51)(6.981,-6.000){2}{\rule{1.450pt}{1.200pt}}
\multiput(715.00,410.26)(0.738,-0.505){4}{\rule{2.357pt}{0.122pt}}
\multiput(715.00,410.51)(7.108,-7.000){2}{\rule{1.179pt}{1.200pt}}
\multiput(727.00,403.26)(0.835,-0.505){4}{\rule{2.529pt}{0.122pt}}
\multiput(727.00,403.51)(7.752,-7.000){2}{\rule{1.264pt}{1.200pt}}
\multiput(740.00,396.26)(0.835,-0.505){4}{\rule{2.529pt}{0.122pt}}
\multiput(740.00,396.51)(7.752,-7.000){2}{\rule{1.264pt}{1.200pt}}
\multiput(753.00,389.25)(0.962,-0.509){2}{\rule{2.900pt}{0.123pt}}
\multiput(753.00,389.51)(6.981,-6.000){2}{\rule{1.450pt}{1.200pt}}
\multiput(766.00,383.26)(0.738,-0.505){4}{\rule{2.357pt}{0.122pt}}
\multiput(766.00,383.51)(7.108,-7.000){2}{\rule{1.179pt}{1.200pt}}
\multiput(778.00,376.25)(0.962,-0.509){2}{\rule{2.900pt}{0.123pt}}
\multiput(778.00,376.51)(6.981,-6.000){2}{\rule{1.450pt}{1.200pt}}
\multiput(791.00,370.26)(0.835,-0.505){4}{\rule{2.529pt}{0.122pt}}
\multiput(791.00,370.51)(7.752,-7.000){2}{\rule{1.264pt}{1.200pt}}
\multiput(804.00,363.25)(0.792,-0.509){2}{\rule{2.700pt}{0.123pt}}
\multiput(804.00,363.51)(6.396,-6.000){2}{\rule{1.350pt}{1.200pt}}
\multiput(816.00,357.26)(0.835,-0.505){4}{\rule{2.529pt}{0.122pt}}
\multiput(816.00,357.51)(7.752,-7.000){2}{\rule{1.264pt}{1.200pt}}
\multiput(829.00,350.25)(0.962,-0.509){2}{\rule{2.900pt}{0.123pt}}
\multiput(829.00,350.51)(6.981,-6.000){2}{\rule{1.450pt}{1.200pt}}
\multiput(842.00,344.25)(0.792,-0.509){2}{\rule{2.700pt}{0.123pt}}
\multiput(842.00,344.51)(6.396,-6.000){2}{\rule{1.350pt}{1.200pt}}
\multiput(854.00,338.26)(0.835,-0.505){4}{\rule{2.529pt}{0.122pt}}
\multiput(854.00,338.51)(7.752,-7.000){2}{\rule{1.264pt}{1.200pt}}
\multiput(867.00,331.25)(0.962,-0.509){2}{\rule{2.900pt}{0.123pt}}
\multiput(867.00,331.51)(6.981,-6.000){2}{\rule{1.450pt}{1.200pt}}
\multiput(880.00,325.25)(0.962,-0.509){2}{\rule{2.900pt}{0.123pt}}
\multiput(880.00,325.51)(6.981,-6.000){2}{\rule{1.450pt}{1.200pt}}
\multiput(893.00,319.25)(0.792,-0.509){2}{\rule{2.700pt}{0.123pt}}
\multiput(893.00,319.51)(6.396,-6.000){2}{\rule{1.350pt}{1.200pt}}
\multiput(905.00,313.25)(0.962,-0.509){2}{\rule{2.900pt}{0.123pt}}
\multiput(905.00,313.51)(6.981,-6.000){2}{\rule{1.450pt}{1.200pt}}
\put(918,305.01){\rule{3.132pt}{1.200pt}}
\multiput(918.00,307.51)(6.500,-5.000){2}{\rule{1.566pt}{1.200pt}}
\multiput(931.00,302.25)(0.792,-0.509){2}{\rule{2.700pt}{0.123pt}}
\multiput(931.00,302.51)(6.396,-6.000){2}{\rule{1.350pt}{1.200pt}}
\multiput(943.00,296.25)(0.962,-0.509){2}{\rule{2.900pt}{0.123pt}}
\multiput(943.00,296.51)(6.981,-6.000){2}{\rule{1.450pt}{1.200pt}}
\put(956,288.01){\rule{3.132pt}{1.200pt}}
\multiput(956.00,290.51)(6.500,-5.000){2}{\rule{1.566pt}{1.200pt}}
\multiput(969.00,285.25)(0.962,-0.509){2}{\rule{2.900pt}{0.123pt}}
\multiput(969.00,285.51)(6.981,-6.000){2}{\rule{1.450pt}{1.200pt}}
\put(982,277.01){\rule{2.891pt}{1.200pt}}
\multiput(982.00,279.51)(6.000,-5.000){2}{\rule{1.445pt}{1.200pt}}
\put(994,272.01){\rule{3.132pt}{1.200pt}}
\multiput(994.00,274.51)(6.500,-5.000){2}{\rule{1.566pt}{1.200pt}}
\put(1007,267.01){\rule{3.132pt}{1.200pt}}
\multiput(1007.00,269.51)(6.500,-5.000){2}{\rule{1.566pt}{1.200pt}}
\put(1020,262.01){\rule{2.891pt}{1.200pt}}
\multiput(1020.00,264.51)(6.000,-5.000){2}{\rule{1.445pt}{1.200pt}}
\put(1032,257.01){\rule{3.132pt}{1.200pt}}
\multiput(1032.00,259.51)(6.500,-5.000){2}{\rule{1.566pt}{1.200pt}}
\put(1045,252.01){\rule{3.132pt}{1.200pt}}
\multiput(1045.00,254.51)(6.500,-5.000){2}{\rule{1.566pt}{1.200pt}}
\put(1058,247.01){\rule{2.891pt}{1.200pt}}
\multiput(1058.00,249.51)(6.000,-5.000){2}{\rule{1.445pt}{1.200pt}}
\put(1070,242.51){\rule{3.132pt}{1.200pt}}
\multiput(1070.00,244.51)(6.500,-4.000){2}{\rule{1.566pt}{1.200pt}}
\put(1083,238.01){\rule{3.132pt}{1.200pt}}
\multiput(1083.00,240.51)(6.500,-5.000){2}{\rule{1.566pt}{1.200pt}}
\put(1096,233.51){\rule{3.132pt}{1.200pt}}
\multiput(1096.00,235.51)(6.500,-4.000){2}{\rule{1.566pt}{1.200pt}}
\put(1109,229.51){\rule{2.891pt}{1.200pt}}
\multiput(1109.00,231.51)(6.000,-4.000){2}{\rule{1.445pt}{1.200pt}}
\put(1121,225.51){\rule{3.132pt}{1.200pt}}
\multiput(1121.00,227.51)(6.500,-4.000){2}{\rule{1.566pt}{1.200pt}}
\put(1134,221.51){\rule{3.132pt}{1.200pt}}
\multiput(1134.00,223.51)(6.500,-4.000){2}{\rule{1.566pt}{1.200pt}}
\put(1147,217.51){\rule{2.891pt}{1.200pt}}
\multiput(1147.00,219.51)(6.000,-4.000){2}{\rule{1.445pt}{1.200pt}}
\put(1159,213.51){\rule{3.132pt}{1.200pt}}
\multiput(1159.00,215.51)(6.500,-4.000){2}{\rule{1.566pt}{1.200pt}}
\put(1172,210.01){\rule{3.132pt}{1.200pt}}
\multiput(1172.00,211.51)(6.500,-3.000){2}{\rule{1.566pt}{1.200pt}}
\put(1185,206.51){\rule{3.132pt}{1.200pt}}
\multiput(1185.00,208.51)(6.500,-4.000){2}{\rule{1.566pt}{1.200pt}}
\put(1198,203.01){\rule{2.891pt}{1.200pt}}
\multiput(1198.00,204.51)(6.000,-3.000){2}{\rule{1.445pt}{1.200pt}}
\put(1210,200.01){\rule{3.132pt}{1.200pt}}
\multiput(1210.00,201.51)(6.500,-3.000){2}{\rule{1.566pt}{1.200pt}}
\put(1223,197.01){\rule{3.132pt}{1.200pt}}
\multiput(1223.00,198.51)(6.500,-3.000){2}{\rule{1.566pt}{1.200pt}}
\put(1236,194.01){\rule{2.891pt}{1.200pt}}
\multiput(1236.00,195.51)(6.000,-3.000){2}{\rule{1.445pt}{1.200pt}}
\put(1248,191.01){\rule{3.132pt}{1.200pt}}
\multiput(1248.00,192.51)(6.500,-3.000){2}{\rule{1.566pt}{1.200pt}}
\put(1261,188.01){\rule{3.132pt}{1.200pt}}
\multiput(1261.00,189.51)(6.500,-3.000){2}{\rule{1.566pt}{1.200pt}}
\put(1274,185.51){\rule{3.132pt}{1.200pt}}
\multiput(1274.00,186.51)(6.500,-2.000){2}{\rule{1.566pt}{1.200pt}}
\put(1287,183.01){\rule{2.891pt}{1.200pt}}
\multiput(1287.00,184.51)(6.000,-3.000){2}{\rule{1.445pt}{1.200pt}}
\put(1299,180.51){\rule{3.132pt}{1.200pt}}
\multiput(1299.00,181.51)(6.500,-2.000){2}{\rule{1.566pt}{1.200pt}}
\put(1312,178.51){\rule{3.132pt}{1.200pt}}
\multiput(1312.00,179.51)(6.500,-2.000){2}{\rule{1.566pt}{1.200pt}}
\put(1325,176.51){\rule{2.891pt}{1.200pt}}
\multiput(1325.00,177.51)(6.000,-2.000){2}{\rule{1.445pt}{1.200pt}}
\put(1337,174.51){\rule{3.132pt}{1.200pt}}
\multiput(1337.00,175.51)(6.500,-2.000){2}{\rule{1.566pt}{1.200pt}}
\put(1350,172.51){\rule{3.132pt}{1.200pt}}
\multiput(1350.00,173.51)(6.500,-2.000){2}{\rule{1.566pt}{1.200pt}}
\put(1363,171.01){\rule{2.891pt}{1.200pt}}
\multiput(1363.00,171.51)(6.000,-1.000){2}{\rule{1.445pt}{1.200pt}}
\put(1375,170.01){\rule{3.132pt}{1.200pt}}
\multiput(1375.00,170.51)(6.500,-1.000){2}{\rule{1.566pt}{1.200pt}}
\put(1388,168.51){\rule{3.132pt}{1.200pt}}
\multiput(1388.00,169.51)(6.500,-2.000){2}{\rule{1.566pt}{1.200pt}}
\put(1401,167.01){\rule{3.132pt}{1.200pt}}
\multiput(1401.00,167.51)(6.500,-1.000){2}{\rule{1.566pt}{1.200pt}}
\put(1414,166.01){\rule{2.891pt}{1.200pt}}
\multiput(1414.00,166.51)(6.000,-1.000){2}{\rule{1.445pt}{1.200pt}}
\put(1426,165.01){\rule{3.132pt}{1.200pt}}
\multiput(1426.00,165.51)(6.500,-1.000){2}{\rule{1.566pt}{1.200pt}}
\sbox{\plotpoint}{\rule[-0.500pt]{1.000pt}{1.000pt}}%
\put(1279,615){\makebox(0,0)[r]{SM}}
\multiput(1299,615)(41.511,0.000){3}{\usebox{\plotpoint}}
\put(1399,615){\usebox{\plotpoint}}
\put(181,627){\usebox{\plotpoint}}
\put(181.00,627.00){\usebox{\plotpoint}}
\put(222.38,623.74){\usebox{\plotpoint}}
\put(263.77,620.48){\usebox{\plotpoint}}
\put(305.15,617.22){\usebox{\plotpoint}}
\put(346.57,614.96){\usebox{\plotpoint}}
\put(387.99,612.69){\usebox{\plotpoint}}
\put(429.40,610.00){\usebox{\plotpoint}}
\put(470.83,608.00){\usebox{\plotpoint}}
\put(512.26,605.90){\usebox{\plotpoint}}
\put(553.69,603.69){\usebox{\plotpoint}}
\put(595.13,602.00){\usebox{\plotpoint}}
\put(636.57,600.12){\usebox{\plotpoint}}
\put(678.03,598.91){\usebox{\plotpoint}}
\put(719.47,597.00){\usebox{\plotpoint}}
\put(760.94,596.00){\usebox{\plotpoint}}
\put(802.41,595.00){\usebox{\plotpoint}}
\put(843.84,593.00){\usebox{\plotpoint}}
\put(885.31,592.00){\usebox{\plotpoint}}
\put(926.78,591.00){\usebox{\plotpoint}}
\put(968.26,590.06){\usebox{\plotpoint}}
\put(1009.73,589.00){\usebox{\plotpoint}}
\put(1051.20,588.00){\usebox{\plotpoint}}
\put(1092.68,587.26){\usebox{\plotpoint}}
\put(1134.15,586.00){\usebox{\plotpoint}}
\put(1175.62,585.00){\usebox{\plotpoint}}
\put(1217.11,584.45){\usebox{\plotpoint}}
\put(1258.57,583.19){\usebox{\plotpoint}}
\put(1300.03,582.00){\usebox{\plotpoint}}
\put(1341.51,581.00){\usebox{\plotpoint}}
\put(1382.96,579.39){\usebox{\plotpoint}}
\put(1424.41,578.00){\usebox{\plotpoint}}
\put(1439,577){\usebox{\plotpoint}}
\end{picture}

%% file: ee_mod.bbl
\begin{thebibliography}{99}

\bibitem{add}
N. Arkani-Hamed, S. Dimopoulos, and G. Dvali, Phys. Lett {\bf B429},
263 (1998);
I. Antoniadis, N. Arkani-Hamed, S. Dimopoulos, and G. Dvali, Phys. Lett.
{\bf B436}, 257 (1998);
N. Arkani-Hamed, S. Dimopoulos, and G. Dvali, Phys. Rev. {\bf D59}, 086004
(1999).

\bibitem{grw}
G.~Giudice, R.~Rattazzi, and J.~Wells, Nucl. Phys. {\bf B544}, 3 (1999)
and revised version 2, e-print hep-ph/9811291.

\bibitem{hlz}
T.~Han, J.D.~Lykken, and R.-J.~Zhang, Phys. Rev. D {\bf 59}, 105006 (1999)
and revised version 4, e-print hep-ph/9811350.

\bibitem{rev}
For an exhaustive list of references on the subject, see: A. Perez-Lorenzana, 
AIP Conf.Proc.{\bf 562} 53 (2001), e-print hep-ph/0008333; 
K. Cheung, eprint hep-ph/0305003.

\bibitem{rs} L. Randall and R. Sundrum, {\it Phys. Rev. Lett.} {\bf 83}
(1999) 3370.

\end{thebibliography}
